\documentclass[12pt,preprint]{aastex}
\usepackage{graphicx}
\usepackage{lscape}
\usepackage{enumerate}

\newcommand{\pivec}{\mbox{\boldmath $\pi$}}
\newcommand{\muvec}{\mbox{\boldmath $\mu$}}
\newcommand{\nuvec}{\mbox{\boldmath $\nu$}}

\newcommand{\gammavec}{\mbox{\boldmath $\gamma$}}
\newcommand{\btheta}{\mbox{\boldmath $\theta$}}
\def\mas{{\rm mas}}
\def\masyr{{\rm mas\,yr^{-1}}}

\lefthead{JUNG ET AL.}
\righthead{ }

\begin{document}
\title{
OGLE-2018-BLG-1269Lb:  A Jovian Planet With A Bright, $I=16$ Host
}

\author{
Youn~Kil~Jung$^{1,28}$, 
Andrew~Gould$^{1,2,3,28}$,
Andrzej~Udalski$^{4,29}$,
Takahiro~Sumi$^{5,30}$,
Jennifer~C.~Yee$^{6,28}$,
Cheongho~Han$^{7,28}$,\\
and \\
Michael~D.~Albrow$^{8}$, Sun-Ju~Chung$^{1,9}$, Kyu-Ha~Hwang$^{1}$, 
Yoon-Hyun~Ryu$^{1}$, In-Gu~Shin$^{1}$, Yossi~Shvartzvald$^{10}$, Wei~Zhu$^{11}$,
Weicheng~Zang$^{12}$, Sang-Mok~Cha$^{1,13}$, Dong-Jin~Kim$^{1}$, Hyoun-Woo~Kim$^{1}$, 
Seung-Lee~Kim$^{1,9}$, Chung-Uk~Lee$^{1,9}$, Dong-Joo~Lee$^{1}$, Yongseok~Lee$^{1,13}$, 
Byeong-Gon~Park$^{1,9}$, Richard~W.~Pogge$^{2}$ \\
(The KMTNet Collaboration) \\
Przemek~Mr{\'o}z$^{4}$, Micha{\l}~K.~Szyma{\'n}ski$^{4}$, Jan~Skowron$^{4}$, Radek~Poleski$^{2,4}$,
Igor~Soszy{\'n}ski$^{4}$, Pawe{\l}~Pietrukowicz$^{4}$, Szymon~Koz{\l}owski$^{4}$, Krzystof~Ulaczyk$^{14}$, 
Krzysztof~A.~Rybicki$^{4}$, Patryk~Iwanek$^{4}$, Marcin~Wrona$^{4}$ \\
(The OGLE Collaboration) \\
Fumio~Abe$^{15}$, Richard~Barry$^{16}$, David~P.~Bennett$^{16,17}$, Ian~A.~Bond$^{18}$, 
Aparna~Bhattacharya$^{16,17}$, Martin~Donachie$^{19}$, Akihiko~Fukui$^{20,21}$, Yuki~Hirao$^{5}$,
Yoshitaka~Itow$^{15}$, Iona~Kondo$^{5}$, Naoki~Koshimoto$^{22,23}$, Man~Cheung~Alex~Li$^{19}$, 
Yutaka~Matsubara$^{15}$, Shota~Miyazaki$^{5}$, Yasushi~Muraki$^{15}$, Masayuki~Nagakane$^{5}$, 
Cl{\'e}ment~Ranc$^{16}$, Nicholas~J.~Rattenbury$^{19}$, Haruno~Suematsu$^{5}$, Denis~J.~Sullivan$^{24}$,
Daisuke~Suzuki$^{25}$, Paul~J.~Tristram$^{26}$, Atsunori~Yonehara$^{27}$ \\
(The MOA Collaboration)
}

\bigskip\bigskip
\affil{$^{1}$Korea Astronomy and Space Science Institute, Daejon 34055, Republic of Korea}
\affil{$^{2}$Department of Astronomy, Ohio State University, 140 W. 18th Ave., Columbus, OH 43210, USA}
\affil{$^{3}$Max-Planck-Institute for Astronomy, K$\rm \ddot{o}$nigstuhl 17, 69117 Heidelberg, Germany}
\affil{$^{4}$Warsaw University Observatory, Al. Ujazdowskie 4, 00-478 Warszawa, Poland}
\affil{$^{5}$Department of Earth and Space Science, Graduate School of Science, Osaka University, Toyonaka, Osaka 560-0043, Japan}
\affil{$^{6}$Center for Astrophysics $|$ Harvard \& Smithsonian, 60 Garden St., Cambridge, MA 02138, USA}
\affil{$^{7}$Department of Physics, Chungbuk National University, Cheongju 28644, Republic of Korea}
\affil{$^{8}$University of Canterbury, Department of Physics and Astronomy, Private Bag 4800, Christchurch 8020, New Zealand}
\affil{$^{9}$University of Science and Technology, Korea, 217 Gajeong-ro Yuseong-gu, Daejeon 34113, Korea}
\affil{$^{10}$IPAC, Mail Code 100-22, Caltech, 1200 E. California Blvd., Pasadena, CA 91125, USA}
\affil{$^{11}$Canadian Institute for Theoretical Astrophysics, University of Toronto, 60 St George Street, Toronto, ON M5S 3H8, Canada}
\affil{$^{12}$Physics Department and Tsinghua Centre for Astrophysics, Tsinghua University, Beijing 100084, China}
\affil{$^{13}$School of Space Research, Kyung Hee University, Yongin 17104, Republic of Korea}
\affil{$^{14}$Department of Physics, University of Warwick, Gibbet Hill Road, Coventry, CV4 7AL, UK}
\affil{$^{15}$Institute for Space-Earth Environmental Research, Nagoya University, Nagoya 464-8601, Japan}
\affil{$^{16}$Code 667, NASA Goddard Space Flight Center, Greenbelt, MD 20771, USA}
\affil{$^{17}$Department of Astronomy, University of Maryland, College Park, MD 20742, USA}
\affil{$^{18}$Institute of Natural and Mathematical Science, Massey University, Auckland 0745, New Zealand}
\affil{$^{19}$Department of Physics, University of Auckland, Private Bag 92019, Auckland, New Zealand}
\affil{$^{20}$Department of Earth and Planetary Science, Graduate School of Science, The University of Tokyo, 7-3-1 Hongo, Bunkyo-ku, Tokyo 113-0033, Japan}
\affil{$^{21}$Instituto de Astrof{\'i}sica de Canarias, V{\'i}a L{\'a}ctea s/n, E-38205 La Laguna, Tenerife, Spain}
\affil{$^{22}$Department of Astronomy, Graduate School of Science, The University of Tokyo, 7-3-1 Hongo, Bunkyo-ku, Tokyo 113-0033, Japan}
\affil{$^{23}$National Astronomical Observatory of Japan, 2-21-1 Osawa, Mitaka, Tokyo 181-8588, Japan}
\affil{$^{24}$School of Chemical and Physical Science, Victoria University, Wellington, New Zealand}
\affil{$^{25}$Institute of Space and Astronautical Science, Japan Aerospace Exploration Agency, Kanagawa 252-5210, Japan}
\affil{$^{26}$University of Canterbury Mt. John Observatory, P.O. Box 56, Lake Tekapo 8770, New Zealand}
\affil{$^{27}$Department of Physics, Faculty of Science, Kyoto Sangyo University, Kyoto 603-8555, Japan}
\footnotetext[28]{The KMTNet Collaboration.}
\footnotetext[29]{The OGLE Collaboration.}
\footnotetext[30]{The MOA Collaboration.}

\begin{abstract}
We report the discovery of a planet in the microlensing event
OGLE-2018-BLG-1269, with planet-host mass ratio 
$q \sim 6\times10^{-4}$, i.e., $0.6$ times smaller than the Jupiter/Sun mass
ratio. Combined with the $Gaia$ parallax and proper motion, 
a strong one-dimensional constraint on the microlens parallax vector 
allows us to significantly reduce the uncertainties of lens physical 
parameters. A Bayesian analysis that ignores any information about 
light from the host yields that the planet is a cold giant 
$(M_{2} = 0.69_{-0.22}^{+0.44}\,M_{\rm J})$ orbiting a Sun-like star 
$(M_{1} = 1.13_{-0.35}^{+0.72}\,M_{\odot})$ at a distance of 
$D_{\rm L} = 2.56_{-0.62}^{+0.92}\,{\rm kpc}$. 
The projected planet-host separation is 
$a_{\perp} = 4.61_{-1.17}^{+1.70}\,{\rm au}$. 
Using {\it Gaia} astrometry, we show that the blended light 
lies $\la 12\,\mas$ from the host and therefore must be either the host star 
or a stellar companion to the host. 
An isochrone analysis favors the former possibility at $>99.6\%$. 
The host is therefore a subgiant. 
For host metallicities in the range of $0.0 \leq {\rm [Fe/H]} \leq +0.3$, 
the host and planet masses are then in the range of 
$1.16 \leq M_{1}/M_{\odot} \leq 1.38$ and 
$0.74 \leq M_{2}/M_{\rm J} \leq 0.89$, respectively. 
Low host metallicities are excluded. 
The brightness and proximity of the lens make the event 
a strong candidate for spectroscopic followup both 
to test the microlensing solution and to further characterize the system.

\end{abstract}

\section{Introduction}

Although microlensing events have been repeatedly observed toward the Galactic bulge field, 
only a few tests of the microlensing solutions have been possible. This is mainly because 
microlensing is an inherently rare phenomenon and the lensing objects are often very faint. 
A microlensing event occurs when two stars at different distances (a foreground lens and 
a background source) are aligned to within of order $1\,$mas along the line of sight. 
This suggests that even in the densest field of the sky (i.e., the Galactic center), 
only about one among a million stars is likely to undergo a microlensing event at a 
given moment \citep{paczynski91,griest91}. In addition, these events are, in most cases, 
not repeating and relatively brief ($t_{\rm E} \sim 20\,$days), where $t_{\rm E}$ 
is the Einstein timescale. While microlensing is sensitive to any lenses distributed along
the Galactocentric distance, the most typical lens stars are $M$ dwarfs because they 
are the most common population of stars in the Galaxy. Hence, the lenses are usually 
very faint (with absolute magnitudes of $M_{I} \sim 8$). Considering that microlensing 
observations are conducted toward crowded fields in which stellar images are severely blended, 
the faintness of the lens makes it challenging to make follow-up observations of the lens 
after the event is over. As a result, there exist only few cases in which the solutions 
for the lenses are checked by follow-up observations.

The most explicit way to check the microlensing solution is to directly observe the lens 
from high resolution imaging. For typical lensing events, the lens proper motion relative 
to the source is $\mu \sim 5~{\rm mas~yr^{-1}}$. This suggests that for direct lens imaging 
with currently available high-resolution instruments, one needs to wait $\sim10-20$ years 
until the lens is separated sufficiently from the source. As a result, this test has been done 
only for a limited number of events \citep{alcock01,kozlowski07,batista15,bennett15,bhattacharya18,vandorou19,bennett20}.

An alternative way to test the microlensing solution regardless of the lens-source motion 
is to conduct spectroscopic observations \citep{han19}. Such an observation may enable one 
to directly measure the lens spectral type from spectroscopic information such as 
temperature, surface gravity, and metallicity. Then, one can check the solution by comparing 
the measured spectral type with the prediction from photometric data. However, this method 
can only be applied provided that the lens is bright enough to be spectroscopically resolved 
at high contrast with the source and unrelated neighbors.

The microlensing solution can be checked by radial velocity (RV) observations \citep{yee16}. 
However, measuring the RV signal for a typical lens is very difficult because of its faintness 
and its slow motion relative to the source. In such conditions, the light from the lens is usually 
contaminated by the blended light, which will significantly dilute the signal from the target of 
interest. For the same reason, the RV observations for stellar lenses with planetary companions 
will be further complicated because their expected radial velocities (${\cal O}~{\rm m~s^{-1}}$) are much 
smaller than those of stellar binaries (${\cal O}~{\rm km~s^{-1}}$). Therefore, the RV measurement on a 
microlensing target also requires a rare lens that is close to us and/or bright enough to be clearly 
visible in the blended light.

In rare cases for which the lens is bright, the microlensing solution can also be checked by 
analyzing the light curve acquired from photometric observations. For solutions with a measured 
lens mass $M$ and a distance $D_{\rm L}$, one can estimate the color and brightness of the lens. 
If these estimates are close to the blended light, it is likely that the lens flux comprises a 
significant portion of the blended flux \citep{han18}. Because the lens is bright, it can be then 
observed in high resolution images as an additional light blended with the source flux. 
Hence, one can check the solution by identifying the lens from the excess flux. For example, 
\citet{bennett10} observed the multiple planetary event OGLE-2006-BLG-109 \citep{gaudi08} 
using the Keck adaptive optics (AO), and confirmed that the light from the lens measured from high 
resolution images is consistent with that predicted from modeling.

Here, we present an analysis of OGLE-2018-BLG-1269. The event was generated by a cold giant planet 
orbiting a Sun-like star, i.e., with the planet-host mass ratio of $q = M_{2}/M_{1} \sim 6\times10^{-4}$. 
The planetary perturbation was densely covered by the Korea Microlensing Telescope Network \citep[KMTNet:][]{kim16}, 
and the parallax and the proper motion of the baseline object was independently measured by $Gaia$. A Bayesian analysis 
suggests that the planet is close to us and the host is associated with the blended light. These make 
the event a strong candidate for high resolution imaging as well as ${\rm 10\,m}$ and ${\rm 30\,m}$ 
spectroscopic observations to test the microlensing solution and 
to further characterize this planetary system.

\section{Observation}

OGLE-2018-BLG-1269, $({\rm RA}, {\rm Dec})_{\rm J2000} = $(17:58:46.42, $-27$:37:04.6) or 
$(l, b) = (2^\circ\hskip-2pt .61,-1^\circ\hskip-2pt .82)$ in Galactic coordinates, 
was first discovered on July 12 by the Optical Gravitational Lensing Experiment 
\citep[OGLE:][]{udalski15} survey and alerted by its Early Warning System \citep{udalski03}. The event 
was in the OGLE BLG504.27 field, with a nominal cadence of ten times per night 
using the ${\rm 1.3\,m}$ Warsaw Telescope located at the Las Campanas Observatory in Chile. 
The apparent $I$-band magnitude of baseline object is $I_{\rm base} \sim 15.8$. We note that 
as will be discussed in Section~\ref{sec:four}, the microlensed source is heavily blended and 
only $\sim4\%$ of the baseline flux comes from the source.

This event was independently found on August 5 by the Microlensing Observations in Astrophysics 
\citep[MOA:][]{sumi03} survey. In the MOA alert system \citep{bond01}, 
and it was listed as MOA-2018-BLG-293. 
The MOA survey monitored the event with a 15 minute cadence using the ${\rm 1.8\,m}$ MOA-II telescope 
located at Mt.\ John Observatory in New Zealand.

The KMTNet survey also discovered the event from its annual post-season analysis \citep{kim18} 
and cataloged it as KMT-2018-BLG-2418. This survey used three ${\rm 1.6\,m}$ telescopes 
that are distributed over three different continents, i.e., Chile (KMTC), South Africa (KMTS), 
and Australia (KMTA). The event was in two offset fields (BLG03 and BLG43), and so was monitored 
with the cadence of four times per hour.

OGLE and KMTNet images were primarily obtained in the $I$ band, while some $V$-band images 
were taken solely to measure the source color. MOA images were obtained in a customized $R$ band, 
which is approximately the sum of the standard $R$ and $I$ band. These data were then reduced 
using pipelines of the survey groups \citep{wozniak00,bond01,albrow09}, 
which are variants of difference image analysis \citep[DIA:][]{tomaney96,alard98}.

\section{Light Curve Analysis}

\begin{figure}[th]
\epsscale{0.9}
\plotone{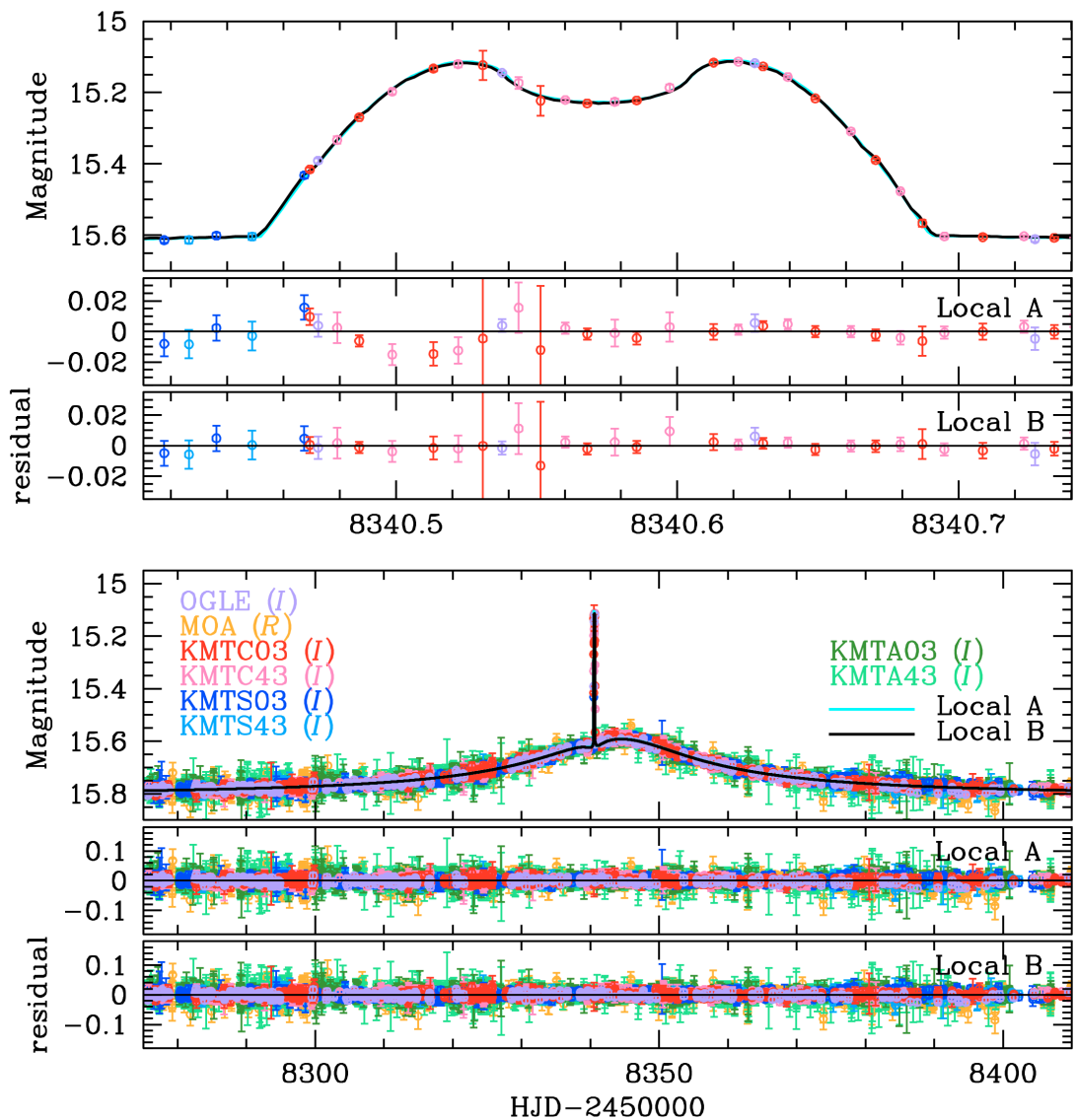}
\caption{\label{fig:one}
Light curve of OGLE-2018-BLG-1269. The upper panel shows 
a zoom of the short-term anomaly centered at ${\rm HJD}' \sim 8340.58$.
The cyan and black curves are the best-fit models from Table 1. 
}
\end{figure}

Figure~\ref{fig:one} shows the light curve of OGLE-2018-BLG-1269. This light 
curve mostly follows a standard \citet{paczynski86} curve except for the 
very short time interval $8340.4 < {\rm HJD}'(={\rm HJD} - 2450000) < 8340.8$, 
during which OGLE and KMTC observations caught a strong anomaly consisting of two 
strong spikes with a U-shaped trough. Such an anomaly typically occurs when a 
source crosses a pair of caustics formed by a binary lens with $q \ll 1$, 
i.e., a planetary system. Hence, we fit the light curve with the binary-lens 
single-source (2L1S) model.

In cases of standard 2L1S models, the lensing magnification, $A(t)$, can be 
described by seven nonlinear parameters. The first three are the geometric 
parameters $(t_{0}, u_{0}, t_{\rm E})$: the time of closest lens-source approach, 
the impact parameter (scaled to the angular Einstein radius $\theta_{\rm E}$), 
and the timescale, respectively. The next three $(s, q, \alpha)$ are the parameters 
that describe the binarity of the lens: the projected companion-host separation 
(scaled to $\theta_{\rm E}$), their mass ratio, and their orientation angle 
(relative to the source trajectory), respectively. The last parameter is the 
source radius $\rho = \theta_{*}/\theta_{\rm E}$, where $\theta_{*}$ is the 
angular source radius..

With these nonlinear parameters, we perform a systematic 2L1S analysis 
by adopting the modeling procedure of \citet{jung15}. We first derive 
initial estimates of $(t_{0}, u_{0}, t_{\rm E})$ by fitting the 
single-lens single-source (1L1S) model to the event with the anomaly 
excluded. We also derive an initial estimate of $\rho = 8\times10^{-4}$ 
based on the source brightness and $t_{\rm E}$ from the 1L1S fit. Next, 
we carry out a dense search over a grid of $s$, $q$, and $\alpha$. For this, 
we divide the parameter space into $200\times200\times21$ grids in the 
range of $-1 < {\rm log}~s < 1$, $-5 < {\rm log}~q < 0$, and $0 < \alpha < 2\pi$, 
respectively. At each $({\rm log}~s, {\rm log}~q, \alpha)$ grid point, 
we fix 
$({\rm log}~s, {\rm log}~q)$ and then fit the light curve by allowing the 
remaining parameters $(t_{0}, u_{0}, t_{\rm E}, \alpha, \rho)$ to vary in 
a Markov Chain Monte Carlo (MCMC).

\begin{figure}[th]
\epsscale{0.9}
\plotone{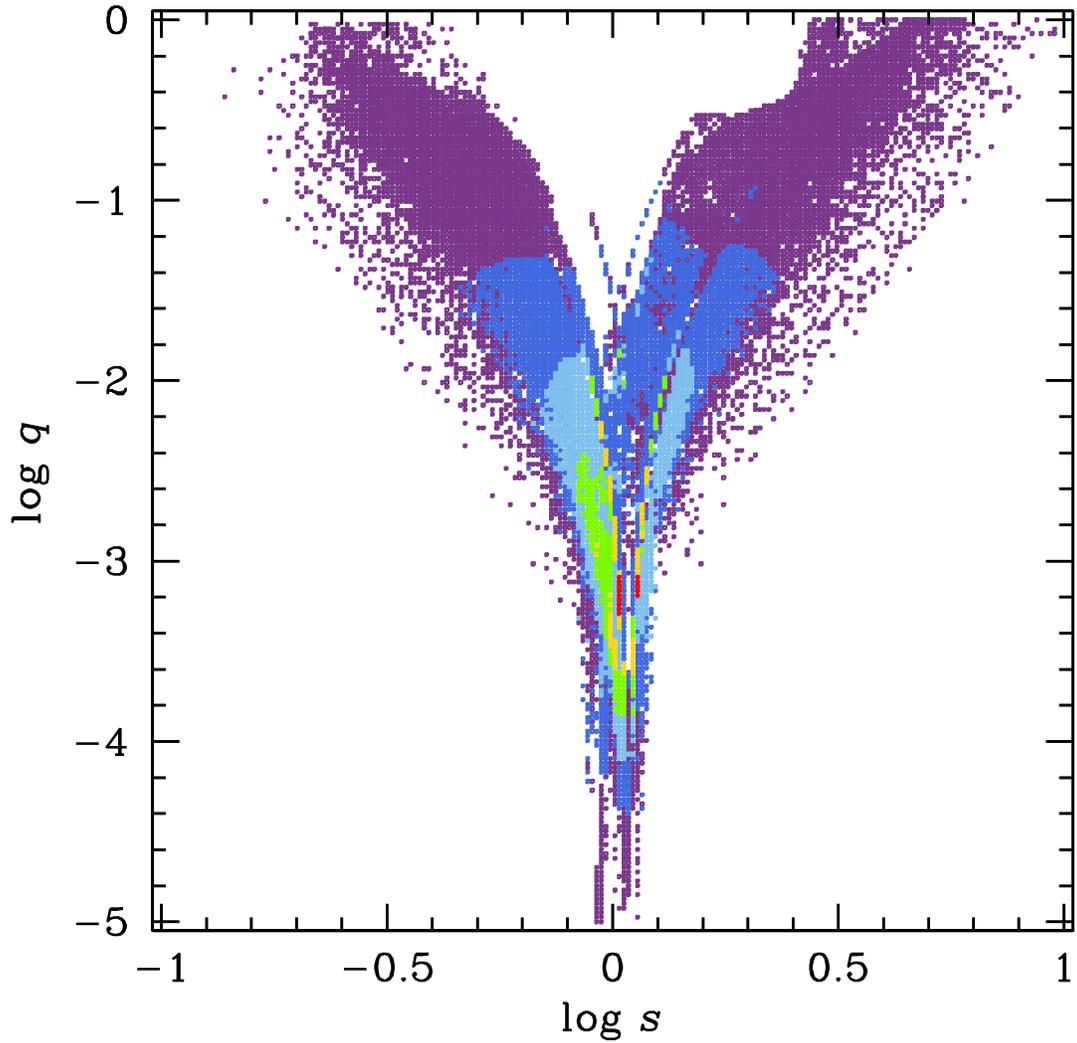}
\caption{\label{fig:two}
$\Delta\chi^{2}$ map in the $({\rm log}\,s, {\rm log}\,q)$ plane 
derived from the grid search. The six colors 
(red, yellow, green, light blue, blue, purple) 
represent the grid with 
$\Delta\chi^2 < [(1\,n)^{2}, (2\,n)^{2}, (3\,n)^{2}, (4\,n)^{2}, (5\,n)^{2}, (6\,n)^{2}]$, 
where $n = 40$.
}
\end{figure}

\begin{figure}[th]
\epsscale{0.9}
\plotone{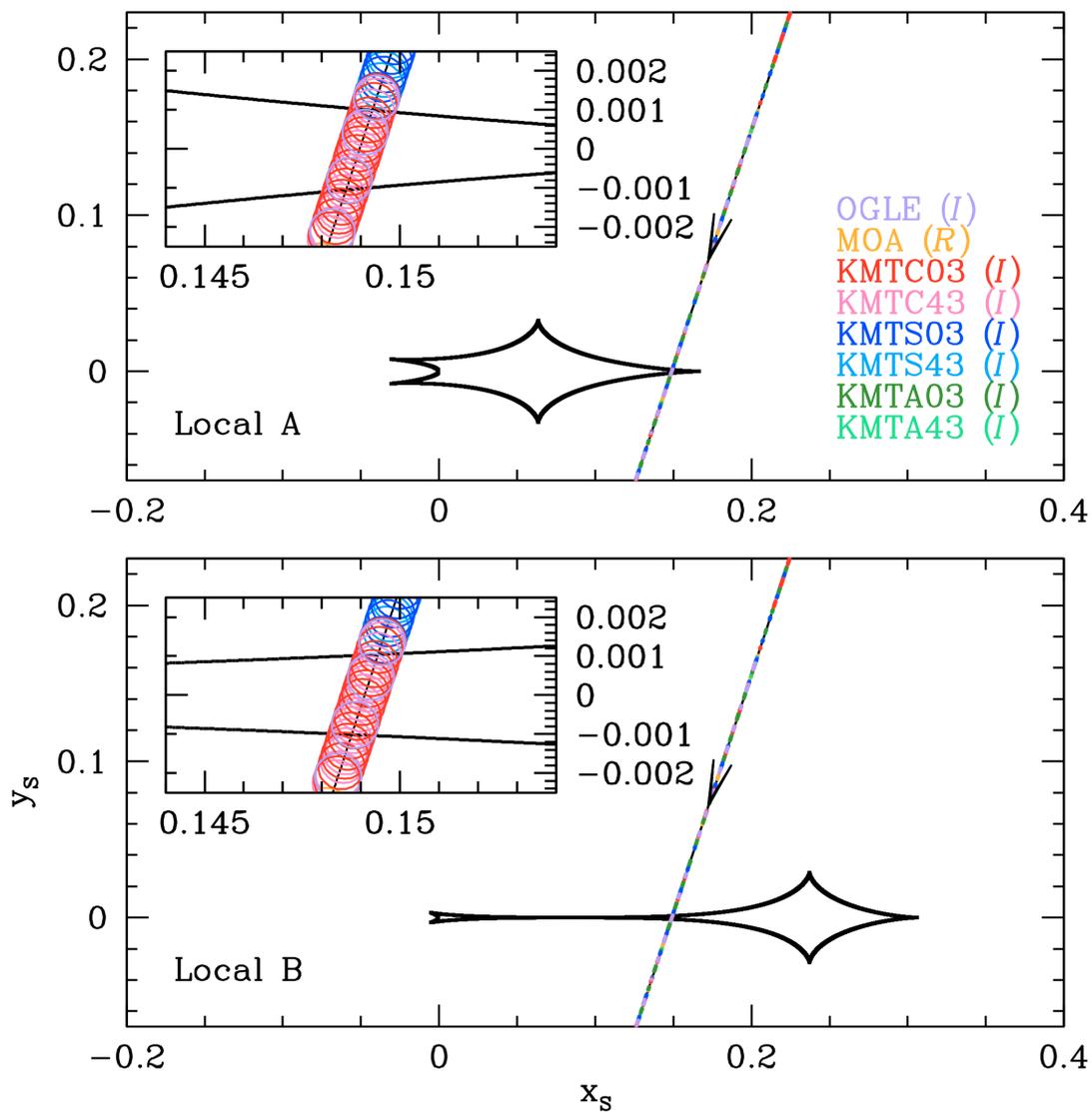}
\caption{\label{fig:three}
Caustic structures for the two solutions. In each panel, 
the black curve is the source trajectory and the open circles 
on the trajectory (scaled by the source radius $\rho$) are the source 
locations at the times of observation. The inset shows 
the zoom of the caustic crossing region. 
}
\end{figure}

We identify two local minima in the resulting $\Delta\chi^{2}$ map 
in the $({\rm log}~s, {\rm log}~q)$ plane (see Figure~\ref{fig:two}). 
We then further refine these minima by optimizing all fitting parameters, 
and finally find that they converge to the two points, 
i.e., $(s, q) = (1.03, 5.94\times10^{-4})$ and $(s, q) = (1.13, 5.93\times10^{-4})$. 
See Table~\ref{table:one}. The MCMC results shows that the best-fit parameters of 
the two solutions are consistent within $1\sigma$ (except for the separation $s$). 
However, the solution ``Local A'' ($s = 1.03$) is disfavored relative to 
the solution ``Local B'' ($s = 1.13$) by $\Delta\chi^{2} = 37$. In addition, 
the former solution has clear systematic residuals in the short-lived anomaly 
region as presented in the upper panel of Figure~\ref{fig:one}. Therefore, 
we exclude the ``Local A'' solution. The caustic structures for the two 
solutions are shown in Figure~\ref{fig:three}.

The timescale of the standard solution ($t_{\rm E} \sim 71~{\rm days}$) 
comprises a substantial portion of Earth's orbit period. Hence, we 
additionally check whether the standard fit further improves by 
introducing the microlens parallax \citep{gould92,gould00}, 
\begin{equation}
\pivec_{\rm E} \equiv \pi_{\rm E}{\muvec_{\rm rel} \over \mu_{\rm rel}};~~~~~\pi_{\rm E} = {\pi_{\rm rel} \over \theta_{\rm E}},
\label{eq1}
\end{equation}
where $(\muvec_{\rm rel}, \pi_{\rm rel})$ are the relative lens-source (geocentric proper motion, parallax).
To account for the parallax, we add two parameters $(\pi_{{\rm E},N}, \pi_{{\rm E},E})$ to the standard model, 
i.e., the north and east component of $\pivec_{\rm E}$ in equatorial coordinates. 
Note that the measurements of $\theta_{\rm E}$ and $\pi_{\rm E}$ allow one to 
determine the lens total mass $M$ and distance $D_{\rm L}$ through the relations 
\begin{equation}
M = {\theta_{\rm E} \over \kappa\pi_{\rm E}};~~~~~D_{\rm L} = {{\rm au} \over \pi_{\rm E}\theta_{\rm E} + \pi_{\rm S}},
\label{eq2}
\end{equation}
where $\kappa = 4G/(c^{2}{\rm au}) \sim 8.14~{\rm mas}~M_{\odot}^{-1}$, $\pi_{\rm S} = {\rm au}/D_{\rm S}$ 
is the source parallax, and $D_{\rm S}$ is the source distance. Then, one can further 
determine the lens physical properties $(M_{1}, M_{2}, a_{\perp})$ from the measured $s$ and $q$, 
where $a_{\perp}=sD_{\rm L}\theta_{\rm E}$ is the physical projected companion-host separation.

The microlens parallax (due to the annual motion of Earth) can be partially mimicked 
by orbital motion of the binary lens \citep{batista11}. This implies that one should 
simultaneously consider the lens orbital motion when incorporating $\pivec_{\rm E}$ 
into the fit. Hence, we also model the orbital effect with two linearized parameters 
$(ds/dt, d\alpha/dt)$, which are the instantaneous change rates of $s$ and $\alpha$, 
respectively \citep{dominik98}.

Based on the ``Local B'' solution, we now fit the light curve with eleven fitting 
parameters $(t_{0}, u_{0}, t_{\rm E}, s, q, \alpha, \rho, \pi_{{\rm E},N}, \pi_{{\rm E},E}, ds/dt, d\alpha/dt)$. 
We also check a pair of solutions with $u_{0} > 0$ and $u_{0} < 0$ to consider the 
ecliptic degeneracy, which takes roughly 
$(u_{0}, \alpha, \pi_{{\rm E},N}, d\alpha/dt) \rightarrow -(u_{0}, \alpha, \pi_{{\rm E},N}, d\alpha/dt)$ \citep{skowron11}. 
From this modeling, we find that the two parameters $(ds/dt, d\alpha/dt)$ are weakly constrained. 
We therefore only consider the MCMC trials that satisfy the condition $\beta < 0.8$. Here, $\beta$ 
is the projected kinetic to potential energy ratio \citep{dong09}  
\begin{equation}
\beta \equiv \left({{\rm KE} \over {\rm PE}}\right)_{\perp} = {{\kappa}M_{\odot}{\rm yr}^{2} \over 8\pi^{2}}{\pi_{\rm E} \over \theta_{\rm E}}{s^{3}{\gamma}^{2} \over (\pi_{\rm E} + \pi_{\rm S}/\theta_{\rm E})^{3}};~~~~~
\gammavec \equiv [(ds/dt)/s, d\alpha/dt],
\label{eq3}
\end{equation}
where we adopt $\pi_{\rm S} = 0.13\pm0.01\,$mas based on the distance 
to the giant clump in the event direction \citep{nataf13}.

The results are listed in Table~\ref{table:one}. We find that the addition of 
higher-order effects does not significantly improve the fit, which only provides 
$\Delta\chi^{2} \sim 8$. This implies that it is difficult to characterize the 
lens system from the fitted parallax parameters alone. Hence, we make a Bayesian 
analysis with Galactic model priors to constrain the lens physical parameters. 
Nevertheless, despite the low level of fit improvement, the analysis gives 
a strong one-dimensional (1-D) constraint on the parallax vector $\pivec_{\rm E}$ 
as seen in Figure~\ref{fig:four}. The short direction of these contours 
corresponds to the direction of Earth's instantaneous acceleration at $t_0$, 
namely $\psi=266.7^\circ$ (north through east), which induces an approximately 
antisymmetric distortion on the light curve around $t_0$. In addition, the seven 
standard parameters are comparable between all the solutions (including the 
standard solution), with the exception of the sign of $u_{0}$. 
Therefore, we take the measured microlens parallax into consideration 
in our Bayesian analysis (e.g., \citealt{jung19}).

\section{Physical Parameter Estimates}
\label{sec:four}

\subsection{Color-Magnitude Diagram (CMD)}
\label{sec:cmd}

The normalized source radius $\rho$ is precisely measured (see Table~\ref{table:one}). 
This implies that we can measure $\theta_{\rm E} = \theta_{*}/\rho$ 
provided that we can estimate the angular source radius $\theta_{*}$. 
The Einstein radius is related to the lens mass $M$ and the relative 
parallax $\pi_{\rm rel}$ by
\begin{equation}
\theta_{\rm E} \equiv \sqrt{{\kappa}M{\pi_{\rm rel}}};~~~~~\pi_{\rm rel} = {\rm au}\left({1 \over D_{\rm L}} - {1 \over D_{\rm S}}\right).   
\label{eq4}
\end{equation}
Then, we can use the measured $\theta_{\rm E}$ to constrain the lens properties. 
Hence, we first estimate $\theta_{*}$ by following the approach of \citet{yoo04}.

\begin{figure}[th]
\epsscale{0.9}
\plotone{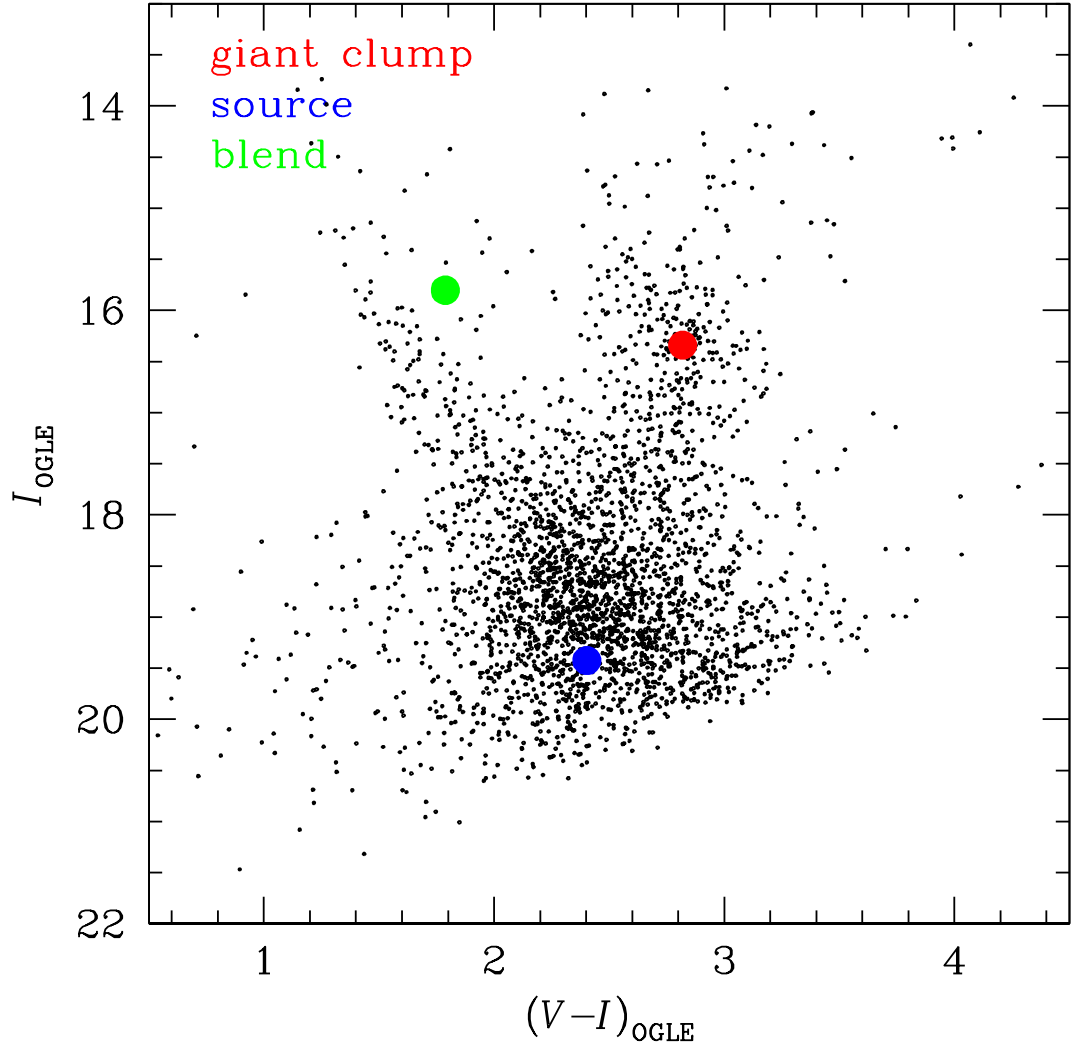}
\caption{\label{fig:five}
Color-magnitude diagram for stars around OGLE-2018-BLG-1269 
obtained from the KMTC03 pyDIA reduction calibrated to OGLE-III photometry. 
The locations of the microlensed source, the centroid of giant clump (GC), 
and the blended light are marked by blue, red, and green circles, respectively. 
}
\end{figure}

Based on the KMTC03 pyDIA reduction calibrated to the OGLE-III 
catalog \citep{szymanski11}, we build a $(V-I, I)$ color-magnitude 
diagram (CMD) with stars centered on the event location (see Figure~\ref{fig:five}). 
We next find the source position of $(V-I, I)_{\rm S} = (2.40\pm0.02, 19.42\pm0.01)$ 
from the best-fit model. We also estimate the giant clump (GC) centroid as 
$(V-I, I)_{\rm GC} = (2.82\pm0.05, 16.34\pm0.07)$, which yields an offset 
\begin{equation}
\Delta(V-I, I) = (V-I, I)_{\rm GC} - (V-I, I)_{0,{\rm GC}} = (1.76\pm0.05, 1.98\pm0.07),
\label{eq5}
\end{equation}
where $(V-I, I)_{0,{\rm GC}} = (1.06, 14.36)$ is the intrinsic GC centroid \citep{bensby13,nataf13}. 
Using this offset, we obtain the dereddened source position as 
$(V-I, I)_{0,{\rm S}} = (V-I, I)_{\rm S} - \Delta(V-I, I) = (0.64\pm0.05, 17.44\pm0.07)$. 
This suggests that the source is either a late F or an early G dwarf.

We then apply $(V-I, I)_{0,{\rm S}}$ to the $VIK$ relation \citep{bessell88} and 
$(V-K)/\theta_{*}$ relation \citep{kervella04} to derive 
\begin{equation}
\theta_{*} = 0.948\pm0.068~\mu{\rm as},
\label{eq6}
\end{equation}
where we add $5\%$ error in quadrature to $\theta_{*}$ to account for the 
uncertainty of $(V-I, I)_{0,{\rm GC}}$ and the color/surface-brightness 
conversion of the Galactic-bulge population relative to locally calibrated stars. 
With the measured $\rho$, we obtain 
\begin{equation}
\theta_{\rm E} = 1.602\pm0.118~{\rm mas}.
\label{eq7}
\end{equation}
The geocentric relative lens-source proper motion is then
\begin{equation}
\mu_{\rm rel} = {\theta_{\rm E} \over t_{\rm E}} = 8.29\pm0.61~{\rm mas}~{\rm yr^{-1}}.
\label{eq8}
\end{equation}

The unusually large values of $\theta_{\rm E}$ and $\mu_{\rm rel}$ suggest that 
the lens lies in the Galactic disk. From the definition of $\theta_{\rm E}$ (Equation~(\ref{eq4})),
\begin{equation}
\pi_{\rm rel} = 0.22~{\rm mas}\biggl({\theta_{\rm E}\over 1.6\,{\rm mas}}\biggr)^2
\biggl({M\over 1.4\,M_\odot}\biggr)^{-1}.
\label{eq9}
\end{equation} 
Thus, given the lens flux constraint, that will be discussed in the following subsection, 
the lens must be $D_{\rm L} = {\rm au}/(\pi_{\rm rel} + \pi_{\rm S}) \la 3~{\rm kpc}$ 
(unless the lens is black hole). We note that the measured $\mu_{\rm rel}$ is also 
consistent with the typical values of disk lenses.

\subsection{{\it Gaia} PPPM of Baseline Object}
\label{sec:GaiaPPPM}

{\it Gaia} data \citep{gaia16,gaia18,luri18} 
will play a critical role in the derivation of 
the lens physical characteristics, in several different respects. 
As has become relatively common in recent years, we will make use of the 
{\it Gaia} proper-motion measurement of the ``baseline object''. However, 
in contrast to most events with such a measurement, in this case, 
the baseline object is strongly dominated by the lens (or at least a 
stellar component of the lens system). Thus, the {\it Gaia} parallax 
measurement is also relevant\footnote{By contrast, {\it Gaia} parallax 
measurements of microlens sources, which are nearly all in the bulge, 
are of no practical use.}. Moreover, we will in this paper, for the 
first time, make use of the {\it Gaia} position measurement of the baseline object. 
That is, we will use the full position, parallax, proper motion (PPPM) {\it Gaia} 
solution at various points in the analysis. Hence, we introduce all of these 
measurements here, together with some context and cautions on their use.

{\it Gaia} reports PPPM values (at epoch 2015.5) of \hfil\break\noindent 
(R.A.,Decl.)$_{\rm J2000}$ = (17:58:46.4171136073, $-27$:37:04.543560775) 
$\pm (0.16,0.14)\,\mas$,
\begin{equation}
\pi_{\rm G}= 0.73\pm 0.18\,\mas,
\label{eqn:pigaia}
\end{equation}
and
\begin{equation}
\muvec_{\rm G}(N, E) = (-1.24\pm0.26, -1.58 \pm 0.31)~{\rm mas}\,{\rm yr^{-1}}. 
\label{eqn:mugaia}
\end{equation} 
{\it Gaia} also reports all $10$ correlation coefficients, but the 
only one of interest for our purposes is the one associated with the 
last equation, $0.31$. 
Before continuing, we note that {\it Gaia} parallaxes have a color-dependent 
zero-point error. For relatively red stars (due to intrinsic color or reddening), 
the shift is measured to be $\pi_{\rm shift} = 0.055\,{\rm mas}$ \citep{zinn19}. 
Hence, we correct the baseline-object parallax to be
\begin{equation}
\pi_{\rm base} = 0.78\pm0.18\,\mas.
\label{eqn:pibase}
\end{equation}

While the {\it Gaia} PPPM catalog is by far the best large-scale 
astrometric database ever constructed, its performance in the crowded 
fields of the Galactic bulge is not at the same level as in high-latitude 
fields, nor even as in the other parts of the Galactic plane. 
For example, \citet{hirao20} found that the {\it Gaia} proper-motion 
measurement of the baseline-object of OGLE-2017-BLG-0406 was spurious. 
This itself shows that {\it Gaia} measurements in crowded fields 
must be treated with caution.

However, \citet{hirao20} also showed, based on generally more 
precise (and likely, more accurate) OGLE proper motions of 
stars in the same field, that the reported {\it Gaia} proper motions of 
most stars are very reliable. In particular, after \citet{hirao20} 
eliminated stars with $\sigma(\pi)/\pi<-2$ 
(which included OGLE-2017-BLG-0406S itself) and those with 
$\sigma(\mu_{\rm north}) > 0.6\,\masyr$ or $\sigma(\mu_{\rm east})> 0.6\,\masyr$, 
that only 1--2\% of {\it Gaia} proper motions were $>3\,\sigma$ outliers. 
However, the {\it Gaia} proper-motion errors had to be renormalized 
by a factor $2.2$ to enforce $\chi^2$/dof = $1$. Although the exact 
reason for this renormalization is not known, it is likely that 
bulge-field crowding is a major contributing cause. In particular, 
the {\it Gaia} mirror has an axis ratio of three, meaning that the 
{\it Gaia} point spread function (PSF) has the same ratio. Hence, 
as {\it Gaia} observes a field at various random orientations, 
light from faint ambient stars can enter the elongated PSF ``aperture'', 
leading to ``random'' shifts in the astrometric centroid. This 
can lead to ``excess noise'' relative to the photon-based error 
estimates, and this ``excess noise'' is tabulated as the 
``astrometric excess noise sig (AENS)'' parameter. In so far as 
this ``excess noise'' is truly ``random'', it just degrades the 
measurement, which is reflected in the reported uncertainties. 
However, because it is likely due to real stars, whose positions 
change very little, and because the observing pattern is also 
not truly ``random'', this ``excess noise'' can lead to systematic 
errors that are larger than the random errors.

In their study of {\it Gaia} proper-motion errors, \citet{hirao20} 
considered stars with AENS$<10$, so their results strictly apply 
to such stars. They did not notice any trends in behavior with AENS, 
and (though not specifically reported), they also did not notice 
any trends for AENS at a few times this level.

For OGLE-2018-BLG-1269, {\it Gaia} reports AENS=10.4. We therefore 
apply the \citet{hirao20} error renormalization to the above {\it Gaia} 
measurements. Although \citet{hirao20} only studied proper-motion errors 
(because this is the only quantity for which OGLE measurements 
are superior to {\it Gaia}), we apply this renormalization to all PPPM
quantities.

For the position measurement, the formal errors $(\sim 0.15\,\mas)$ are 
so small that they play no practical role, even after renormalization. 
So we ignore these errors. 
For the parallax measurement, the renormalized error is $\sim 45\%$ 
of the measured value. Hence, its role is mainly qualitative 
confirmation that the lens is nearby. 
The renormalized proper-motion errors ($\sim 0.7\,\masyr$) 
are still relatively small, and this measurement will play 
a crucial role at several points.

\subsection{Blend is Due to Host and/or Its Companion}
\label{sec:blend_is_lens}

\subsubsection{{\it Gaia} Baseline Object Is $<20\,\mas$ From Source}
\label{sec:gaiabase}

{\it Gaia} astrometry is generally given for epoch 2015.50, whereas 
the event peaked at 2018.61. In order to compare the position of 
the source (at 2018.61) with the position of the {\it Gaia} baseline 
object at the same time, we first propagate the positions of all {\it Gaia} 
stars (including the baseline object) forward in time by 
$\Delta t = 2018.61 - 2015.50 = 3.11\,$yr. 
That is, for each star in the field, $i$, we calculate, 
\begin{equation}
\btheta^{2018.61}_{i,Gaia} = \btheta^{2015.5}_{i,Gaia} + \muvec_{i,Gaia}\Delta{t},
\label{eq10}
\end{equation}
where $\muvec_{i,Gaia}$ is the proper motion of {\it Gaia} object. 
We then cross-match the {\it Gaia} and KMTNet 
pyDIA catalogs within a $2^\prime$ square, excluding entries that fail 
a relatively forgiving cut on an empirical $(G-I)/(V-I)$ color-color 
relation, and with a $1^{\prime\prime}$ astrometric cut (to allow for 
optical distortion of the KMTNet camera). Next, we fit for a 
transformation from {\it Gaia} to KMTNet pyDIA coordinates using all 
matches obtained from the previous step, except the ``baseline object'', 
by minimizing the unrenormalized $\chi^2$, 
\begin{equation}
\chi^2 = \sum_i [\btheta^{2018.61}_{i,\rm KMTNet} -
T_n(\btheta^{2018.61}_{i,Gaia})]^2,
\label{eq11}
\end{equation} 
where $T_n$ is a $n$-th order polynomial transformation, i.e., 
$6$, $12$, and $20$ parameters for $n=(1, 2, 3)$.

We find that the results vary very little but the $n=3$ polynomial fit 
is slightly better than the others. We recursively eliminate outliers, 
of which there are $46$ objects\footnote{We note that roughly half are 
false matches due to the relatively loose matching criteria, and the 
remainder are likely due to corrupted astrometry from unresolved objects.} 
for 487 original matches. The scatter is $19\,$mas, which 
is almost an order of magnitude larger than the typical formal 
propagated errors in $\btheta^{2018.61}_{i,Gaia}$. Although the $Gaia$ 
errors are probably somewhat underestimated in crowded fields \citep{hirao20}, 
it is still the case that this scatter is completely 
dominated by the errors of KMTNet pyDIA astrometry.

We then apply the resulting transformation to the propagated $Gaia$ 
blend position $\btheta^{2018.61}_{{\rm b},Gaia}$, and subtract this 
from the pyDIA source position that is derived from difference images: 
\begin{equation}
\Delta \btheta(N,E) =
\btheta^{2018.61}_{{\rm S}, {\rm KMTNet}}-T_3(\btheta^{2018.61}_{{\rm b}, Gaia}) = 
(6.9, 9.2)\pm(4.4,4.4)\,{\rm mas}.
\label{eq12}
\end{equation}
The error comes from three sources added in quadrature: error in the 
transformation coefficients ($1.0\,$mas), error in the propagated 
value of $\btheta^{2018.61}_{{\rm b},Gaia}$ ($2.0\,$mas), and error in 
the pyDIA measurement of $\btheta^{2018.61}_{{\rm S},{\rm KMTNet}}$ 
($4.0\,$mas).

That is, the {\it Gaia} baseline object lies $11.5\pm 4.4$ mas from 
the source at the time of the event. We repeat this exercise using 
OGLE data and obtain $18.5\pm 4.4$ mas, which confirms that the 
source and the {\it Gaia} baseline object are very close.

\subsubsection{Probability of Chance Superposition is $p=3\times 10^{-6}$}
\label{sec:psuperpose}

Figure~\ref{fig:five} shows that the blend is bright and belongs to 
the foreground main-sequence branch, i.e., 
$(V-I, I)_{\rm b} = (1.79\pm0.02, 15.80\pm0.01)$. 
The surface density of such ``bright'' $(I < 16)$, ``blue'' $(V-I < 2)$ 
foreground stars is only $n = 9\,{\rm arcmin}^{-2}$. 
In the previous subsection, we showed that the {\it Gaia} baseline object is 
within $\delta\theta = 20\,$mas. Therefore, the probability of an unrelated 
bright foreground star lying within $\delta\theta$ of this foreground-star lens is 
$\pi(\delta\theta)^2 n = 3 \times 10^{-6}$. 
Hence, the blend is almost certainly associated with the event, 
rather than being a random interloper.

\subsubsection{Blended Light is Due to the Lens System}
\label{sec:lenssystem}

That is, there are broadly five classes of objects that 
could contribute significantly to baseline object: 
(0) the source, 
(1) a stellar companion to the source, 
(2) the lens host, 
(3) a stellar companion to the host, 
(4) an unrelated ambient star.

Of course, the source does contribute, but this contribution is 
well determined from the microlensing fit, and in the present case 
is also quite small. The remaining four possibilities are candidates 
for the remaining light, i.e., the blend. The $p = 3\times10^{-6}$ 
probability just calculated implies that (4) is ruled out. 
Moreover (1) is also ruled out by the color ($1$ mag bluer than the clump) 
and magnitude ($0.5$ mag brighter than the clump). To be a companion 
of the source (and hence in the bulge) this would have to be a late B dwarf, 
of which there are essentially none in the bulge 
(apart from the star-forming regions near the Galactic Center).

Thus, the blended light is due to either the host, a companion to the 
host, or possibly a combination of the two.  For any of these 
possibilities, the parallax and proper motion of the host are essentially 
equal the parallax and proper motion of the blend because the host 
and it putative companion are at essentially the same distance, 
and their orbital motion is very slow compared the lens-source 
relative motion. Hence, the {\it Gaia} measurements of these quantities 
will act as strong constraints on the estimates of 
the physical parameters of the lens system.

\subsection{Bayesian Analysis}
\label{sec:bayes}

For the Bayesian analysis, we will incorporate the {\it Gaia} astrometric 
measurements in addition to the usual microlensing-parameter measurements.

\subsubsection{Inputs From {\it Gaia}}
\label{sec:gaiainput}

To do so, we first note that for the three parameters ${\bf X} = (\pi,\muvec)$ 
measured by {\it Gaia}, the observed (``baseline object'') quantities 
are related to the underlying physical (``source'' and ``lens'') 
quantities\footnote{In the more general case, one would write 
``source'' and ``blend''. However, in Section~\ref{sec:lenssystem}, 
we established that the blend is the lens (although up to this point 
it is not yet clear whether it can be identified with the host, 
its stellar companion, or both).} by
\citep{kb181292}
\begin{equation}
{\bf X}_{\rm base} = (1-\eta){\bf X}_{\rm L} + \eta{\bf X}_{\rm S},
\label{eqn:splitparms}
\end{equation} 
where $\eta = f_{\rm S}/f_{\rm base}$ is the flux fraction of the 
source in the {\it Gaia} band and $(f_{\rm S}, f_{\rm base})$ are the flux 
of the source and the baseline object, respectively. We estimate 
$\eta$ by noting that the peak of the {\it Gaia} passband is broadly 
consistent with that of the $V$ band and the typical photometric error 
of {\it Gaia} observation is $2\%$. We thereby estimate $\eta = 0.02$ 
based on our result that the blend is $4.2$ mag brighter than the 
source in the $V$ band.

To find the lens parallax $\pi_{\rm L}$, we adopt 
$\pi_{\rm S} =0.13\pm0.01\,\mas$ from \citet{nataf13} and 
we renormalize the errors (by a factor $2.2$) in Equation~(\ref{eqn:pibase}) 
to obtain $\pi_{\rm base}=0.78\pm 0.40\,\mas$. 
We then apply Equation~(\ref{eqn:splitparms}) to $\pi_{\rm base}$ and find
\begin{equation}
\pi_{\rm L} = 0.80\pm0.40\,\mas.
\label{eqn:piL}
\end{equation}

The situation is substantially more complicated for the proper motion. 
First, the microlensing solution gives the amplitude of the lens-source 
relative proper motion in the geocentric frame, 
but the {\it Gaia} proper motion is in the heliocentric frame. 
We can relate these by
\begin{equation}
\muvec_{\rm rel, hel} \equiv \muvec_{\rm L, hel} - \muvec_{\rm S, hel};
\qquad
\muvec_{\rm rel, hel} = \muvec_{\rm rel} + {\pi_{\rm rel} \over {\rm au}}\nuvec_{\oplus,\perp},
\label{eqn:murelhel}
\end{equation}
where $\nuvec_{\oplus,\perp}(N, E) = (-1.7, 18.4)~{\rm km}~{\rm s^{-1}}$ 
is the projected velocity of Earth at $t_0$ and
$(\muvec_{\rm L, hel}, \muvec_{\rm S, hel})$ are the heliocentric
proper motions of the lens and the source, respectively.

In principle, we could fully incorporate 
Equations~(\ref{eqn:splitparms}) and (\ref{eqn:murelhel}) 
into the Bayesian analysis below. However, as we now show, 
for the case of OGLE-2018-BLG-1269 the lens proper motion 
is well approximated by $\muvec_{\rm L,hel} = \muvec_{\rm base}$.

First, we combine 
Equations~(\ref{eqn:splitparms}) and (\ref{eqn:murelhel}) to yield
\begin{equation}
\muvec_{\rm L,hel} = \muvec_{\rm base} + \eta[\muvec_{\rm rel} + {\pi_{\rm rel} \over {\rm au}}\nuvec_{\oplus,\perp}],
\qquad
\muvec_{\rm S,hel} = \muvec_{\rm base} - (1-\eta)[\muvec_{\rm rel} + {\pi_{\rm rel} \over {\rm au}}\nuvec_{\oplus,\perp}].
\label{eqn:muLmuS}
\end{equation} 
Next, we note that for typical final values of $\pi_{\rm rel} = 0.4\,\mas$, 
we have $(\pi_{\rm rel}/{\rm au})\nu_{\oplus,\perp} \simeq 1.6\,{\rm mas\,yr^{-1}}\ll \mu_{\rm rel}$, 
and hence, regardless of the direction of $\muvec_{\rm rel}$, we have 
$\mu_{\rm rel,hel}\simeq\mu_{\rm rel}\simeq 8\,{\rm mas\,yr^{-1}}$. 
Therefore, the second term in the first entry of Equation~(\ref{eqn:muLmuS}) 
is of order $\eta\mu_{\rm rel}\sim 0.16\,{\rm mas\,yr^{-1}}$, which is a factor 
of about four smaller than the renormalized errors in $\muvec_{\rm base}$ ($\sim 0.7\,\masyr$). 
Hence, we adopt $\muvec_{\rm L,hel} = \muvec_{\rm base}$. Then, after renormalizing 
the errors (by a factor $2.2$) in Equation~(\ref{eqn:mugaia}) 
and rotating to Galactic coordinates, we obtain
\begin{equation}
\muvec_{\rm L}(l, b) = (-1.86\pm0.68, 0.75\pm0.57)~{\rm mas}~{\rm yr^{-1}}.
\label{eqn:muL}
\end{equation}

\subsubsection{Bayesian Formalism}
\label{sec:bayesform}

With the four measured constraints $(t_{\rm E}, \theta_{\rm E}, \pi_{\rm L}, \pivec_{\rm E})$, 
we now make a Bayesian analysis following the procedure of \citet{jung18}. 
We first build a Galactic model with models of the mass function (MF), 
density profile (DP), and velocity distribution (VD) of astronomical objects. 
For the MF and DP, we adopt the models used in \citet{jung18}. 
For the VD, we use the proper motion distribution of stars measured by $Gaia$. 
For the source proper motion, we examine a $Gaia$ CMD using red giant stars 
within $3\,{\rm arcmin}$ centered on the event direction and find their mean 
proper motion and standard deviation in Galactic coordinates 
\begin{equation}
\muvec_{\rm S}(l, b) = (-5.93\pm3.10, 0.03\pm2.72)~{\rm mas}~{\rm yr^{-1}}.
\label{eq17}
\end{equation} 
For the lens proper motion, we employ Equation~(\ref{eqn:muL}).

For each solution of $u_{0} > 0$ and $u_{0} < 0$, we draw one billion random events 
based on the adopted Galactic model. For each random event $i$, we then infer the 
four parameters $(t_{\rm E}, \theta_{\rm E}, \pi_{\rm L}, \pivec_{\rm E})_{i}$ and 
find the $\chi^{2}$ difference between the inferred and the measured values, i.e., 
\begin{equation}
\chi_{{\rm gal}, i}^{2} = \chi_{i}^{2}(t_{\rm E}) + \chi_{i}^{2}(\theta_{\rm E}) + \chi_{i}^{2}(\pi_{\rm L}) + \chi_{p, i}^{2};~~~~~
\chi_{p, i}^{2} = \sum(a_i - a_0)_{j}c_{jk}^{-1}(a_i - a_0)_{k}, 
\label{eq19}
\end{equation}
where $\mathbf{a}_{i} = \pivec_{{\rm E}, i} = (\pi_{{\rm E},N},~\pi_{{\rm E},E})_{i}$ 
is the inferred parallax, and $\mathbf{a}_{0}$ and $c_{jk}$ are the measured $\pivec_{\rm E}$ 
and its covariance matrix, respectively. We next find the likelihood of the event by 
${\rm P}_{i} = {\rm exp}(-\chi_{{\rm gal}, i}^{2}/2)\times\Gamma_{i}$, 
where $\Gamma_{i} \propto \theta_{{\rm E},i}\mu_{{\rm rel},i}$ is the lensing event rate.

For each random event $i$, we also infer the lens position in the calibrated CMD, 
i.e., $(V-I, I)_{\rm L}$, in order to check whether the lens flux predicted from 
the Bayesian estimates is consistent with the blended flux. For this, we first 
construct a set of isochrones with different metallicities and ages \citep{spada17}, 
i.e., with $[{\rm Fe/H}] = (-0.5, 0.0, +0.3)$ and ${\rm age} = (2, 4, 6, 8, 10)$ Gyr. 
In each isochrone $j$, we next estimate the absolute $I$-band magnitude $M_{I,{\rm L},i,j}$ 
and intrinsic $(V-I)_{0,{\rm L},i,j}$ color of the lens from the inferred lens mass $M_{i}$. 
We then find the dereddened lens magnitude in the $I$ and $V$ band by 
$I_{0,{\rm L},i,j}$ = $M_{I,{\rm L},i,j} + 5{\rm log}(D_{{\rm L},i}/{\rm pc}) - 5$ and 
$V_{0,{\rm L},i,j}$ = $I_{0,{\rm L},i,j} + (V-I)_{0,{\rm L},i,j}$. We next estimate the 
extinction to the lens $A_{\lambda,{\rm L},i}$ using the partial extinction model 
\citep{bennett15,beaulieu16},
\begin{equation}
A_{\lambda,{\rm L},i} = {{1 - e^{-\left\vert{D_{{\rm L},i}/\tau_{\rm dust}}\right\vert}} \over {1 - e^{-\left\vert{D_{{\rm S},i}/\tau_{\rm dust}}\right\vert}}}A_{\lambda,{\rm S}},
\label{eq20}
\end{equation}
where the index $\lambda$ denotes the passband and 
$\tau_{\rm dust} = (0.12~{\rm kpc})/{\rm sin}(b)$ is the dust scale height. 
Here $A_{\lambda,{\rm S}}$ is the extinction to the source, for which we adopt 
$A_{I,{\rm S}} = 1.98$ and $E(V-I)_{\rm S} = 1.76$ from our CMD analysis 
(Equation~(\ref{eq5})). We then derive $(V-I, I)_{{\rm L},i,j}$ using $A_{\lambda,{\rm L},i}$, 
and bin the CMD by these lens positions with the likelihood $P_{i}$.

We emphasize that in this initial analysis, we completely ignore 
constraints coming from the blended light.  That is, we neither impose 
any constraint on the lens light (such as not to exceed the blended light) 
nor consider the possibility that the lens is responsible for 
the blended light. At this point, we simply ``predict'' the lens color 
and magnitude based on the $(t_{\rm E}, \theta_{\rm E}, \pi_{\rm L}, \pivec_{\rm E})$ 
[or $(t_{\rm E}, \theta_{\rm E}, \pi_{\rm L})$] constraints together with 
the Galactic model and model isochrones. We investigate the role of the 
blended light in constraining the solution only after comparing these 
predictions to the observed blended light.

\subsubsection{Bayesian Results}
\label{sec:bayesresult}

We finally investigate the posterior probabilities of the lens properties 
from all random events. We note that to check the contribution of the 
$\pivec_{\rm E}$ constraint on the Bayesian estimates, we also explore 
the posterior probabilities with $(t_{\rm E}, \theta_{\rm E}, \pi_{\rm L})$ 
constraints.

\begin{figure}[th]
\epsscale{0.9}
\plotone{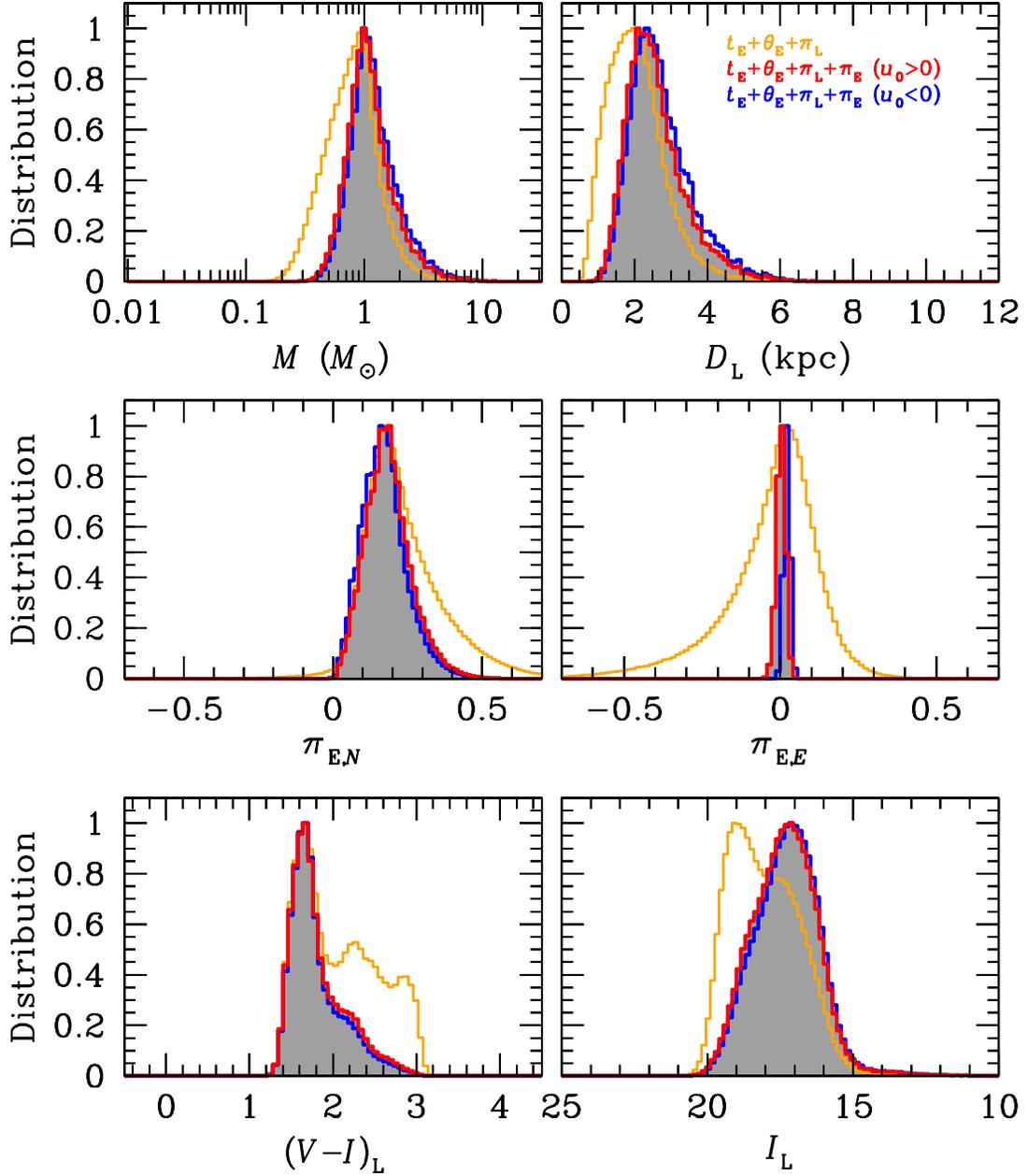}
\caption{\label{fig:six}
Posterior distributions for the lens parameters. In each panel, 
the yellow curve shows the distribution obtained from the timescale $t_{\rm E}$, 
the angular Einstein radius $\theta_{\rm E}$, and the lens parallax $\pi_{\rm L}$ constraints.
The blue and red curves are, respectively, the distributions for the $u_{0} > 0$ and $u_{0} < 0$ 
solutions derived from $t_{\rm E}$, $\theta_{\rm E}$, $\pi_{\rm L}$, and the microlens parallax $\pivec_{\rm E}$ constraints.
}
\end{figure}

The results from the constraints $(t_{\rm E}, \theta_{\rm E}, \pi_{\rm L}, \pivec_{\rm E})$ 
are shown in Figure~\ref{fig:six} and listed in Table~\ref{table:two}. 
Also listed are the total Galactic-model probability $P_{\rm tot} = \sum{P_{i}}$ 
and the net relative probability $P_{\rm net} = P_{\rm tot}P_{\rm lc}$, 
where $P_{\rm lc} = {\rm exp}(-\Delta\chi^{2}/2)$ is the relative fit probability and 
$\Delta\chi^{2}$is the $\chi^{2}$ difference between the two solutions. Here, 
$\phi_{\rm hel} = {\rm tan^{-1}}\,[\mu_{{\rm rel,hel}}(E)/\mu_{{\rm rel,hel}}(N)]$ 
is the orientation angle of $\muvec_{{\rm rel,hel}}$.

We find that the measured $\pivec_{\rm E}$ from modeling gives a 
strong constraint on the probability distributions. However, we also 
find that the host-mass ranges of the two solutions 
($u_0 > 0$ and $u_0 < 0$) are somewhat different from each other. 
Because $\pivec_{\rm E}$ is the only prior constraint that differs 
significantly between the two solutions and because $\pivec_{\rm E}$ 
is connected to the lens mass (Equation~(\ref{eq2})), the difference 
would imply that the Galactic-model priors disfavor one of the 
solutions. To check this, we also draw the two-dimensional ($2$-D) 
likelihood ${\cal L}$ for the lens parameters obtained from the 
$(t_{\rm E}, \theta_{\rm E}, \pi_{\rm L})$ and the 
$(t_{\rm E}, \theta_{\rm E}, \pi_{\rm L}, \pivec_{\rm E})$ constraints. 
See Figure~\ref{fig:seven}. We note that the black and grey error bars 
in the three $(\pi_{{\rm E}, N}, \pi_{{\rm E}, E})$ planes are the errors 
of $\pivec_{\rm E}$ listed in Table~\ref{table:one} for the $u_0 > 0$ 
and $u_0 < 0$ solutions, respectively.  From this figure, we find that 
the measured $\pivec_{\rm E}$ from both $u_{0} > 0$ and $u_0<0$ solutions 
are consistent at the $1\,\sigma$ level. 
This mutual consistency is reflected in the almost equal values of 
$P_{\rm tot}$ (ratio $0.92$) in Table~\ref{table:two}. Thus, our 
estimates should reflect the weighted average of the two solutions, 
although these hardly differ.

Hence, these results, which do not yet incorporate constraints from the 
blended light, suggests that the host is a Sun-like star with 
$M_{1} = 1.13_{-0.35}^{+0.72}~M_{\odot}$ located at a distance of 
$D_{\rm L} = 2.56_{-0.62}^{+0.92}~{\rm kpc}$. Then, the microlensing 
companion is a planet with $M_{2} = 0.69_{-0.22}^{+0.44}~M_{\rm J}$ 
separated (in projection) from the host by 
$a_{\perp} = 4.61_{-1.17}^{+1.70}~{\rm au}$. 
That is, the planet is a cold giant lying beyond the snow line, i.e., 
$a_{sl} = 2.7 {\rm au}(M_{1}/M_{\odot}) \sim 3.1~{\rm au}$.

\begin{figure}[th]
\epsscale{0.9}
\plotone{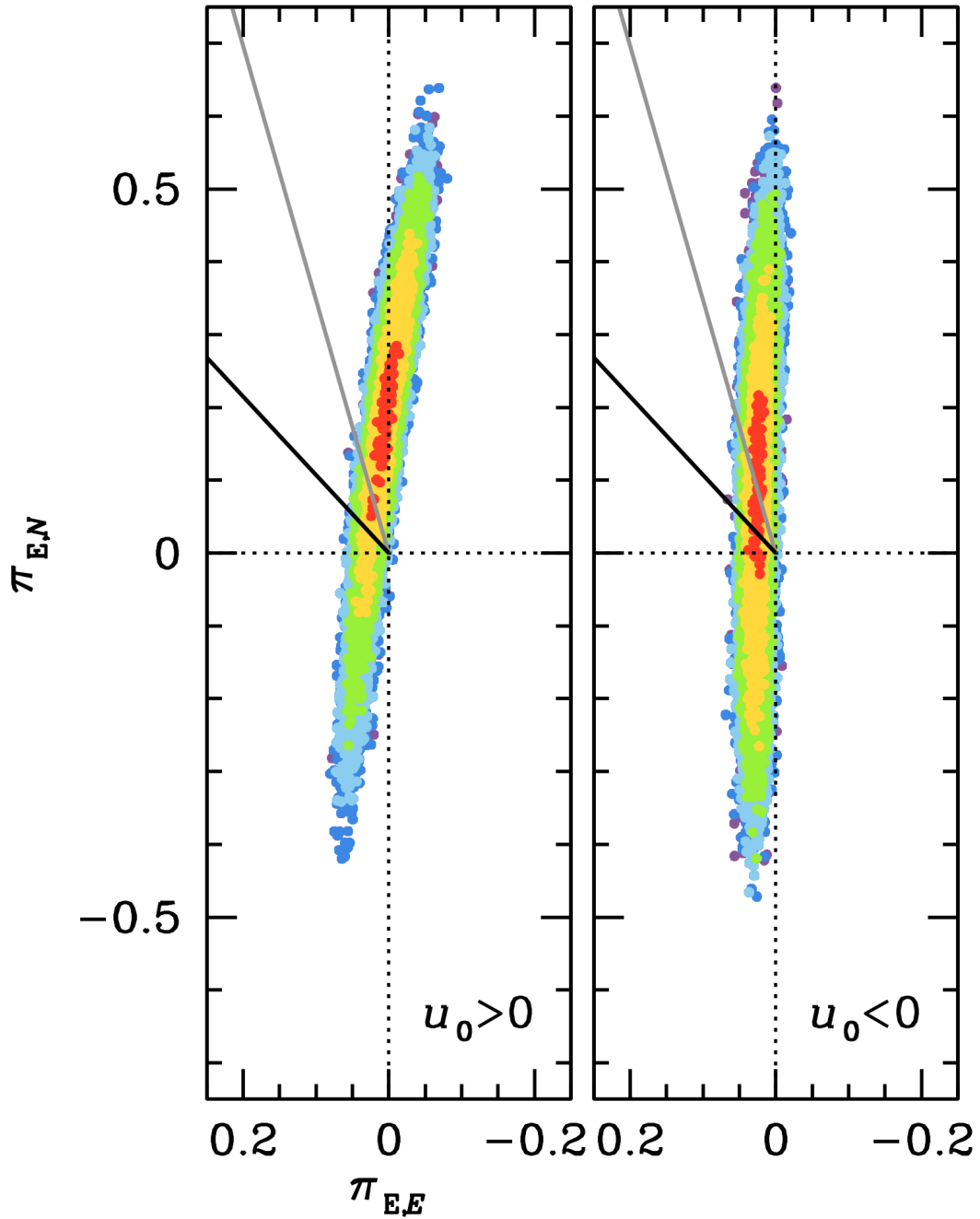}
\caption{\label{fig:four}
$\Delta\chi^{2}$ maps in $(\pi_{{\rm E}, N}, \pi_{{\rm E}, E})$ 
plane obtained from the two solutions ($u_{0} > 0$ and $u_{0} < 0$). 
Except that $n = 1$, the color notation is identical 
to that of Figure~\ref{fig:two}.  The two rays in each panel at
$\phi=16^\circ$ (gray) and $\phi=43^\circ$ (black) represent 
the $1\,\sigma$ range of the direction of the lens-source relative
motion that is derived within the Bayesian analysis.  See the final
paragraph of Section~\ref{sec:four}.  It is the imposition of this
constraint on the 1-D parallax contours in this figure that
forces the two solutions to have very similar and relatively small
mass ranges.
}
\end{figure}

\begin{figure}[th]
\epsscale{0.9}
\plotone{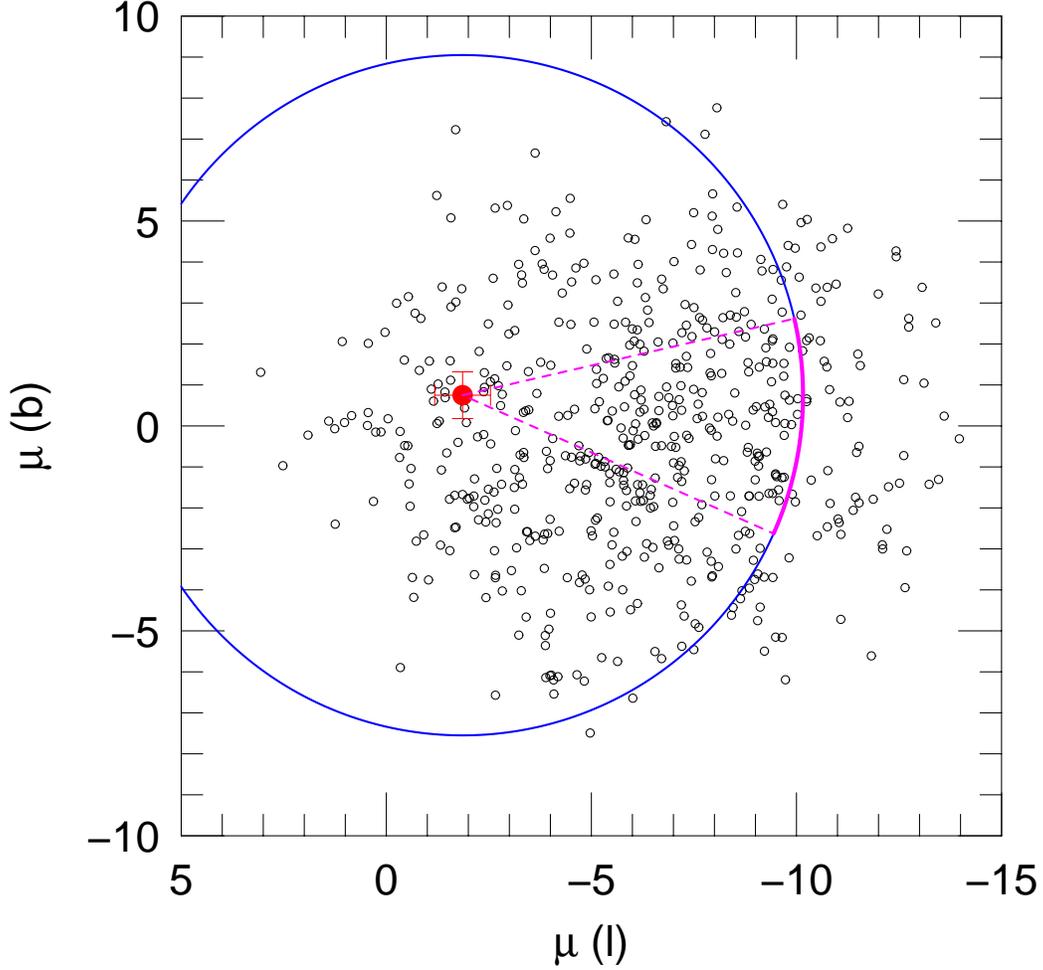}
\caption{\label{fig:eight}
The proper motion of the lens (red) is shown relative to those
of bulge clump giant stars (black) within a $3^\prime$ circle
centered of OGLE-2018-BLG-1269, which are a tracer of the
general population of bulge sources.  The blue circle is the locus
of possible source proper motions, given that
$\mu_{\rm rel} = |\muvec_{\rm L}-\muvec_{\rm S}| = 8.3\pm 0.6\,{\rm mas\,yr^{-1}}$
The magenta arc is the portion of this circle that is consistent
at the $1\,\sigma$ level with the proper-motion distribution of bulge
sources in the $b$ direction.  When this arc is projected onto the
microlens contours (Figure~\ref{fig:four}), it strongly constrains
the parallax along the long direction of those contours.
}
\end{figure}

It is of some interest to understand how the Bayesian analysis constrains 
the lens mass to a range of a factor $\sim 2.4$ at the $1\,\sigma$ level 
despite the fact that $\pi_{\rm E}$ varies by a factor $10$ at $1\,\sigma$ 
(see Figure~\ref{fig:four}), while $M = \theta_{\rm E}/\kappa\pi_{\rm E}$. 
The main reason is that the direction of $\muvec_{\rm rel}$ (and so the 
direction of $\pivec_{\rm E}$) is reasonably well constrained by the 
{\it Gaia} measurement of $\muvec_{\rm L}$ together with the relatively 
large value of $\mu_{\rm rel}\simeq 8.3\,{\rm mas\,yr^{-1}}$. This is 
illustrated in Figure~\ref{fig:eight}, which shows $\muvec_{\rm L}$ 
and a blue circle to represent all possible $\muvec_{\rm S}$ that are 
consistent with $\mu_{\rm rel} = |\muvec_{\rm L} -\muvec_{\rm S}|$. 
The magenta arc of this circle represents the $1\,\sigma$ range of 
$\mu_{\rm S}$ as given by Equation~(\ref{eq17}). 
The arc of allowed (at $1\,\sigma$) directions is shown by dashed lines. 
This same arc, rotated to Equatorial coordinates 
(and displayed as lens-source rather than source-lens motion) 
is shown in Figure~\ref{fig:four}, with boundaries $\phi=16^\circ$ 
to $\phi=43^\circ$ (north through east). The first point to note is that, 
for both $u_0 > 0$ and $u_0 < 0$, this arc subtends a region that is 
almost entirely contained within the $1\,\sigma$ $\pivec_{\rm E}$ contours, 
implying mutual consistency between two constraints on the direction 
that are entirely independent. Second, for the $u_0 < 0$ contours, 
which are almost perfectly vertical, 
we can evaluate the range of $\pi_{\rm E}$ as 
$\pi_{\rm E}(16^\circ)/\pi_{\rm E}(43^\circ) = \csc(16^\circ)/\csc(43^\circ) = 2.47$. 
This confirms that the relatively tight constraints on $M$ in the Bayesian analysis 
derive from the application of the directional constraint on lens-source relative 
motion (Figure~\ref{fig:eight}) to the $1$-D parallax contours from the light-curve 
analysis (Figure~\ref{fig:four}). That is, the Bayesian mass estimate comes 
primarily from a combination of measured microlensing parameters and measured 
lens proper motion, while the Galactic model enters mainly by constraints 
on the source proper motion.

\section{Blended Light Is Due Mainly to the Host} 
\label{sec:five}

As we discussed in Section~\ref{sec:four}, the facts that the lens host 
is known (from the Bayesian analysis) to be a roughly solar-mass 
foreground-disk star and that there is such a roughly solar-mass 
foreground-disk star projected within $\sim 12\,$mas of the lens 
make it virtually certain that this blended light comes from 
the lens system. That is, the blended light must be due to the host, 
to a companion to the host, or to some combination of the two. 
Here we examine this issue in detail.

\begin{figure}[th]
\epsscale{0.9}
\plotone{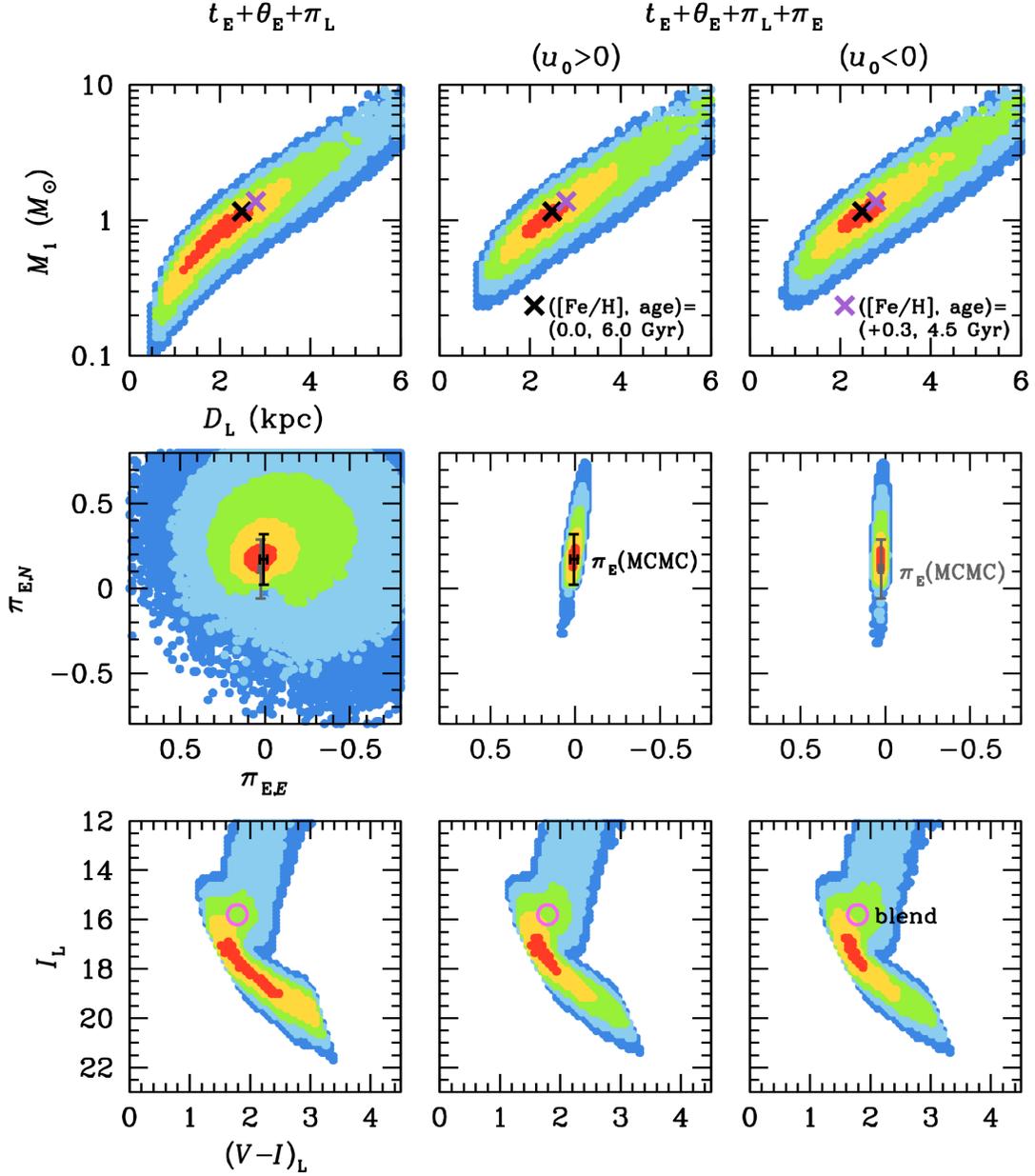}
\caption{\label{fig:seven}
$-2\Delta{\rm ln}{\cal L}$ contours in the $(M_{1}, D_{\rm L})$, $(\pi_{{\rm E}, N}, \pi_{{\rm E}, E})$, and $(V-I, I)_{\rm L}$ planes.
The left three panels show the contours from $(t_{\rm E}$, $\theta_{\rm E}$, $\pi_{\rm L})$ constraints. 
The middle and right three panels show the contours for the $u_0 > 0$ and $u_{0} < 0$ solutions 
from $(t_{\rm E}$, $\theta_{\rm E}$, $\pi_{\rm L}, \pivec_{\rm E})$ constraints. 
The black and grey error bars represent the errors of $\pivec_{\rm E}$ listed in Table~\ref{table:one} 
for the $u_0 > 0$ and $u_0 < 0$ solutions, respectively. 
The magenta circles are the location of the blended light measured from the CMD analysis (see Figure~\ref{fig:five}). 
The black and purple crosses are the blend positions estimated by matching the ``observed'' isochrones 
to the blended light, for the ${\rm [Fe/H]} = 0.0$ and ${\rm [Fe/H]} = +0.3$ isochrones, respectively (see Figure~\ref{fig:nine}). 
The color notation is identical to that of Figure~\ref{fig:four}.
}
\end{figure}

The two lower-right panels of Figure~\ref{fig:seven} show that 
the blend (magenta circle) lies at about $2.5\,\sigma$ from the most 
likely ``prediction'' of the Bayesian analysis 
for both the $u_0 > 0$ and $u_0 < 0$ solutions. However, this nominal 
$2.5\,\sigma$ ``discrepancy'' may simply reflect the fact that stars 
spend far more time on the upper main sequence than they do at the 
location of the blend (i.e., subgiant branch, or possibly end of the turnoff). 
That is, a small range of lens masses from the upper main-sequence is projected
onto a small region of the CMD, but an equally small range of masses 
that are just slightly larger are projected all across the subgiant branch, 
and thus populate the CMD at much lower density. Hence, the blended light 
could be fully consistent with being due to the host, but would show up 
as ``relatively low probability'' on this figure.

We now must take account of the fact that, despite the low prior probability 
that the host is a turnoff/subgiant star (as indicated by it being projected 
against the green contours in Figure~\ref{fig:seven}), there is actually such 
a ``star'' associated with the event, 
i.e., either the host itself, a companion to the host, or a combination of the two. 
We now consider these three possibilities in turn.

\subsection{Blend is Consistent with Being Due to the Host}
\label{sec:five_1}

We first ask whether the host is consistent with being the primary 
contributor to the blended light?  If it is, then there must be a star 
simultaneously consistent with the microlensing properties and the 
blended light. For this analysis, we use the measured Einstein radius 
$\theta_{\rm E}$ and the adopted source parallax 
$\pi_{\rm S}$ (Equation~(\ref{eq2})) to map a set of model isochrones 
to the calibrated CMD. Given $\theta_{\rm E}$ and $\pi_{\rm S}$, 
we can take a star with given mass $M_{\rm iso}$, 
intrinsic color $(V-I)_{0, {\rm iso}}$, and 
absolute magnitude $M_{I, {\rm iso}}$, 
to estimate the distance to the star $D_{\rm iso}$. 
We next find the dereddened $I$- and $V$-band magnitudes by 
$I_{0,{\rm iso}}$ = $M_{I,{\rm iso}} + 5{\rm log}(D_{{\rm iso}}/{\rm pc}) - 5$ 
and $V_{0,{\rm iso}}$ = $I_{0,{\rm iso}} + (V-I)_{0,{\rm iso}}$. 
We then find the position of the star $(V-I, I)_{\rm iso}$ in the 
calibrated CMD using the partial extinction model (Equation~(\ref{eq20})). 
Finally, we build an ``observed'' isochrone with $[M, D, (V-I), I]_{\rm iso}$ 
from all stars listed in the model isochrone. 
For the three cases of ${\rm [Fe/H]} = (-0.5, 0.0, +0.3)$, 
we then construct ``observed'' isochrones with different ages, and 
compare them to the blended light to estimate the blend mass 
$M_{\rm b}$ and distance $D_{\rm b}$.

We find that two ``observed'' isochrones can match the blended light 
(see Figure~\ref{fig:nine}). 
That is, the two curves for $(\rm {[Fe/H]},{\rm age}) = (0.0, 6\,{\rm Gyr})$ 
and $(+0.3, 4.5\,{\rm Gyr})$ pass through the blend position with the offset 
of $(5.5, 8.5)\times10^{-3}$, respectively. 
The estimated mass and distance to the blend are 
$(M_{\rm b}, D_{\rm b}) = (1.16\,M_{\odot}, 2.49\,{\rm kpc})$ for 
$(\rm {[Fe/H]},{\rm age}) = (0.0, 6\,{\rm Gyr})$ and 
$(M_{\rm b}, D_{\rm b}) = (1.38\,M_{\odot}, 2.80\,{\rm kpc})$ for 
$(\rm {[Fe/H]},{\rm age}) = (+0.3, 4.5\,{\rm Gyr})$. 
These estimates imply that for typical disk populations with 
$0 \leq {\rm [Fe/H]} \leq 0.3$, 
$M_{\rm b}$ and $D_{\rm b}$ are in the range of 
$1.16 \leq M_{\rm b}/M_{\odot} \leq 1.38$ and 
$2.49 \leq D_{\rm b}/{\rm kpc} \leq 2.80$, 
respectively. These ranges show remarkable agreement 
with the prediction from the Bayesian analysis (see Figure~\ref{fig:seven}). 
This implies that the host is consistent with causing the blended 
light, which is then a subgiant (or possibly late turnoff) star.

\subsection{Blend as Stellar Companion to the Host (Qualitative Analysis)}
\label{sec:five_2}

We still must consider the possibility that the blended light is primarily 
due to a stellar companion to the host, rather than the host itself. That is, 
it is due to a star that does not directly enter into the microlensing event 
but is gravitationally bound to the host. This alternative explanation for 
the blended light can be conceptually divided into two cases: either the host 
contributes very little light to the blend, or the host and its brighter stellar 
companion both contribute significantly to the observed blended light. When we 
quantitatively evaluate the probability that the host dominates the blended light, 
we will treat these two alternative cases as a single case. However, in the 
qualitative treatment that follows immediately below, we make a conceptual 
distinction between them.

In order to evaluate the three possibilities, i.e., that the blend light
\begin{enumerate}[(1)]
\setlength{\leftskip}{15pt}
\item is dominated by host
\item is dominated by a stellar companion to the host (and the host contributes relatively little light),
\item receives comparable contributions from the host and a stellar companion.
\end{enumerate}
\setlength{\leftskip}{0pt}
\noindent we first divide all single and binary systems containing a subgiant 
(or possibly late turnoff) star into six classes:
\begin{enumerate}[(A)]
\setlength{\leftskip}{15pt}
\item single stars,
\item binaries with orbital periods $P<10^4\,$day 
\item binaries with orbital periods $P>10^{5.3}\,$day 
\item binaries with  $10^4<P/{\rm day}<10^{5.3}$, and mass ratios $Q<0.5$,
\item binaries with  $10^4<P/{\rm day}<10^{5.3}$, and $0.5<Q<0.9$,
\item binaries with  $10^4<P/{\rm day}<10^{5.3}$, and $0.9<Q<1$.
\end{enumerate}
\setlength{\leftskip}{0pt}
\noindent Using the statistics of \citet{duquennoy91} for solar-type stars, 
we estimate relative fractions $(0.36, 0.20, 0.28, 0.10, 0.05, 0.01)$ 
for the classes $(\rm{A, B, C, D, E, F})$, respectively.

These six classes of systems can contribute the three cases of events as follows. 
Class (A) can contribute only to events of case (1). 
Class (B) cannot contribute to any events with a OGLE-208-BLG-1269 type light 
curve because companions in this period range would have given rise to 
recognizable signals in the light curve ($P<10^4\,$days) or would violate 
the $Gaia$-based source-blend separation measurement ($P<10^{5.3}\,$days).

Class (C) is excluded for cases (2) and (3) because the centroid of light 
would be displaced from the the host (and therefore the host) by more than $12$ mas. 
However, it is permitted for case (1) because the light from the companion would not 
significantly displace the light centroid.

Class (D) can contribute to events of case (1) but not of case (2) or (3). 
That is, the light contributed by the stellar companion would not be enough 
to qualitatively alter the photometric appearance of the combined light 
relative to an isolated turnoff/subgiant star, so case (1) is compatible. 
However, the mass of the host for case (2) is too low to be compatible 
with microlensing constraints (see Table~\ref{table:two}). Therefore, case (2) 
is excluded. And case (3) is also excluded because a $Q < 0.5$ companion cannot 
contribute significantly.

Class (E) can contribute to either case (1) or case (2). Because the 
two stars in the lens system must be on the same isochrone, in the 
class (E) mass-ratio range, the more massive star must be above the 
turnoff and the less massive one must be below the turnoff. Hence, 
they differ by at least one magnitude, which implies that they 
contribute substantially differently to the total light of the blend. 
From the lower-right panels of Figure~\ref{fig:seven}, it is clear that 
over most of this mass-ratio range, the lower-mass star would have a similar 
color to the blend. Hence, the brightness of the higher-mass star would be 
reduced by $\sim 0.1$ to $\sim 0.5$ mag, while its color would hardly be 
altered, relative to the blend. Thus, its position on the CMD would be 
essentially the same as that of the blend. In particular, it would be 
projected against the same green contours, and, in fact, slightly closer 
to the yellow contours.

Because class (F) systems can contribute only to case (3), 
we can now qualitatively evaluate the relative likelihood of case (1) 
(blended light from ``turnoff/subgiant star'' is dominated by the 
host, i.e., classes (A), (C), (D), and part of class (E)) 
and case (2) (blended light from ``turnoff/subgiant star'' is 
dominated by a companion to the host, i.e., part of class (E)). 
Then we will return to case (3).

For class (E), in which there are two stars in the system,
i.e., higher- and lower-mass stars, the overall probability 
of lensing is higher than for a single star by $\sqrt{1 + Q}$ 
because there are two well-separated lenses that could give rise 
to the event. And the relative probability of the lower-mass star 
giving rise to the event is $\sqrt{Q}$. Therefore, relative to 
single-host case, the absolute lensing probabilities of two stars 
scale as $1$ and $\sqrt{Q}$, respectively. We can then approximate 
the lower-mass events of class (E) by $\sqrt{Q} \sim \sqrt{0.7} \sim 0.84$. 
Then, the probability for case (2) relative to case (1) can be directly evaluated: 
$p_2/p_1 = (0.05\times0.84)/(0.36+0.28+0.10+0.05) = 0.05$.

Naively, event case (3) appears highly disfavored because only system 
class (F) contributes to it, and this comprises only $1\%$ of all systems. 
In fact, however, this case requires close examination for proper evaluation.

\begin{figure}[th]
\epsscale{0.9}
\plotone{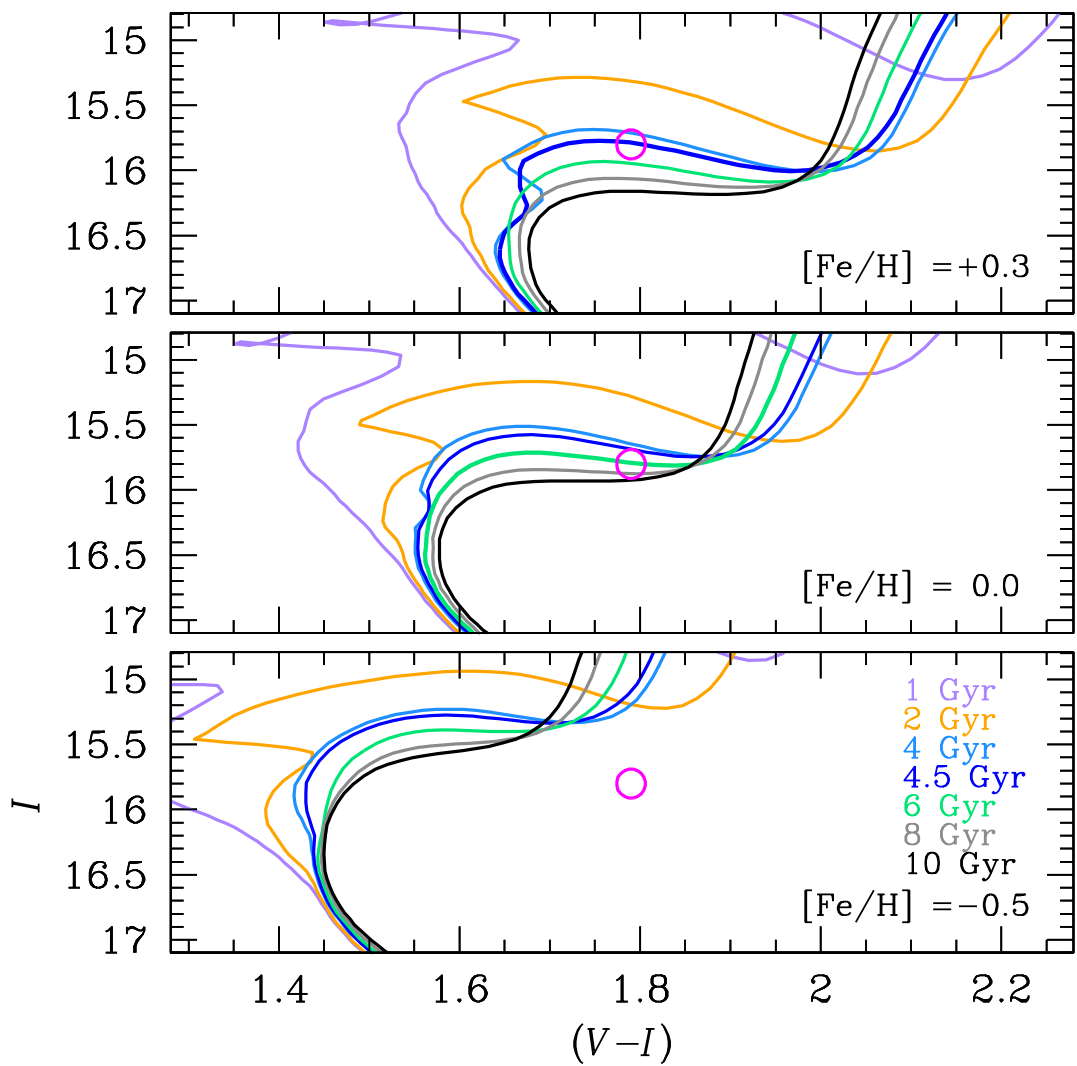}
\caption{\label{fig:nine}
Model isochrones calibrated to the observed CMD. 
In each panel, the curves with different colors are the ``observed'' isochrones 
from different metallicities and ages. The magenta circle is the position of the blended light. 
The green curve $(6\,{\rm Gyr})$ in the middle panel $(\rm {[Fe/H]} = 0.0)$ and the blue curve 
$(4.5\,{\rm Gyr})$ in the upper panel $({\rm [Fe/H]} = +0.3)$ are the two isochrones 
that pass over the observed color and magnitude of the blended light.
}
\end{figure}

We first consider the very special subcase that the host and its companion 
are identical. Then their colors would be the same as the blend, but 
their magnitudes would be $0.75$ mag fainter than the blend. In principle, 
this might have put them on the main sequence. In this case, 
the low relative probability of such binary systems ($1\%$) would 
have been counterbalanced by the fact that main-sequence stars are far 
more common than turnoff/subgiants of the same color. In fact, however, 
Figure~\ref{fig:nine} shows that this position ($0.75$ mag below the blend) 
is not inhabited by stars on any or the fairly broad range of isochrones 
that we have displayed.

If we consider the broader case of approximately (rather than exactly) 
equal masses for the two components ($Q \sim 0.9$), we see that 
essentially the same (above) argument applies to the case. The less 
massive star will be fainter and bluer than the blend, while the more 
massive star will be fainter and redder. The upper panel of 
Figure~\ref{fig:nine} shows that at ${\rm [Fe/H]} = +0.3$, it is 
possible for a star to exist on, e.g., the $10$-Gyr isochrone that is 
$0.3$ mag fainter and somewhat redder than the blend. However, this 
position invalidates the main advantage of event cases that was just 
mentioned above: the lens (or its companion) remains a subgiant and is 
not on the more populous main sequence.  Hence, the probability of this 
solution is very low. Using the same procedure as above, we derive 
$p_3/p_1 = (0.01\times1.9)/(0.36+0.28+0.10+0.05) = 0.02$.

\subsection{Blend as Lens Companion (Quantitative Analysis)}
\label{sec:five_3}

We now conduct a quantitative analysis aimed at both testing the 
qualitative ideas presented above and deriving a more precise quantitative 
result. Our starting point is to draw random events from the same Galactic 
model used for the Bayesian analysis described above and to weight each event 
by the same $(t_{\rm E}, \theta_{\rm E}, \pi_{\rm L}, \pivec_{\rm E})$ priors. 
However, for each simulated event, we either ``accept'' or ``reject'' it 
according to whether the combined light from the host
and {\it some} companion 
drawn from the same isochrone is compatible with the blended light. That is,
each simulated event has a corresponding $I$-band magnitude and $V-I$
color; if there exists a companion along the same isochrone for which
the combined light is compatible with the blend, the event is ``accepted''. 
The entire isochrone is reddened in the same manner as was done for the 
case that the blend is dominated by the host light. We consider the same 
$(3\times 5 = 15)$ isochrones that were analyzed for the host=blend case, 
i.e., case (1). We note that after investigating
these separate-isochrone cases, 
we must still combine them to obtain an overall relative probability of 
case (1) versus cases (2)+(3). This step will also require incorporating 
information about binary frequency.

\begin{figure}[th]
\epsscale{0.9}
\plotone{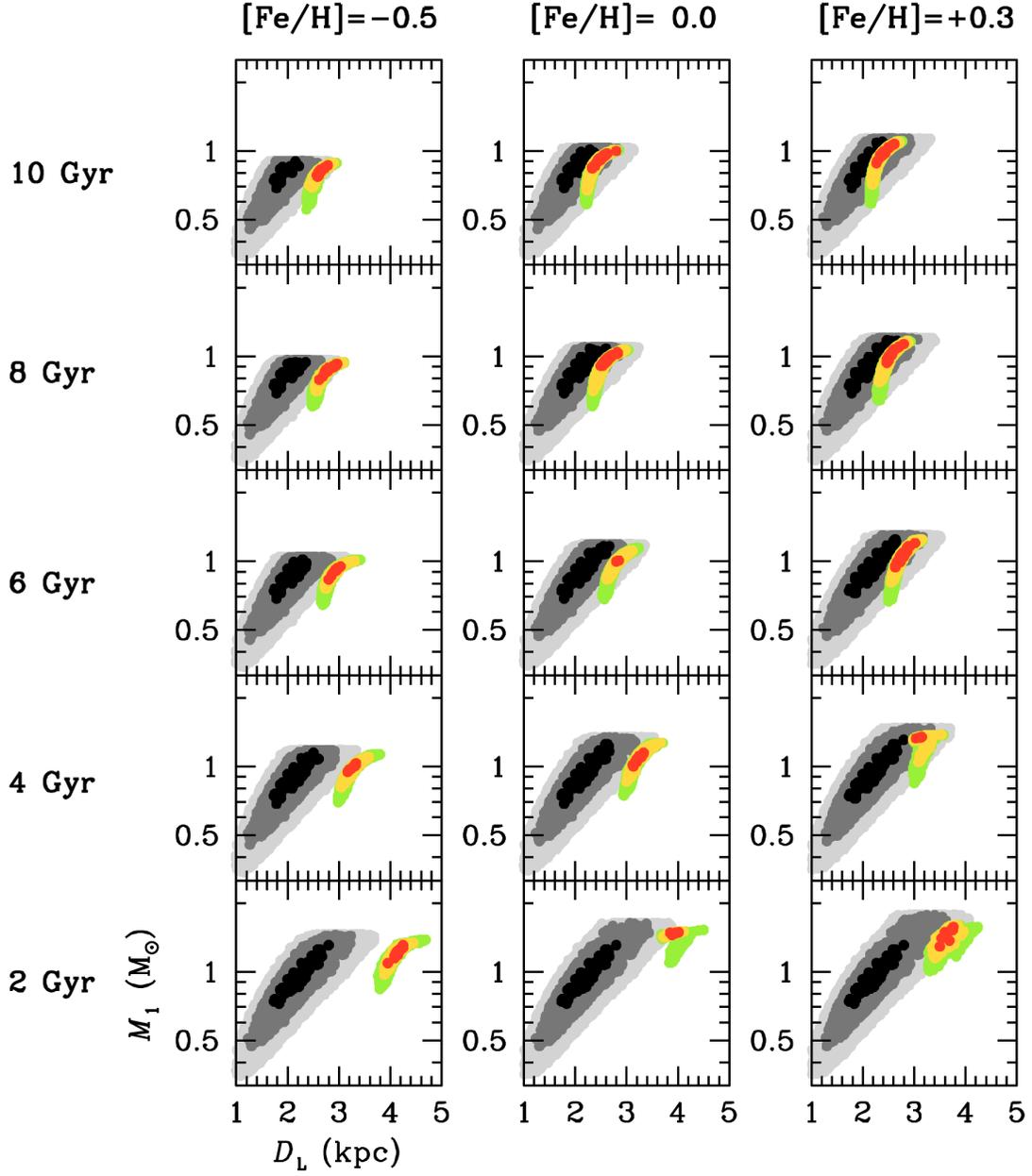}
\caption{\label{fig:ten}
$-2\Delta{\rm ln}{\cal L}$ contours in the $(M_{1}, D_{\rm L})$ plane 
for the [accepted+rejected] and [accepted-only] simulated events based on $15$ model isochrones, 
i.e., ${\rm [Fe/H]} = (-0.5, 0.0, +0.3)$ and ${\rm age} = (2, 4, 6, 8, 10)\,{\rm Gyr}$. 
In each panel, the (black, dark grey, grey) colors are the $(1, 2, 3) \sigma$ contours 
from both ``accepted'' and ``rejected'' events. 
The (red, yellow, green) colors are the $(1, 2, 3) \sigma$ contours from ``accepted'' events.
}
\end{figure}

Figure~\ref{fig:ten} shows separate $(1, 2, 3) \sigma$ contours, 
in the lens mass-distance plane, for the all [accepted+rejected] 
(black, dark grey, grey) and [accepted-only] (red, yellow, green) 
simulated events. We first focus on the five ${\rm [Fe/H]} = -0.5$ 
isochrones. These show that the accepted contours lie well away from 
the contours for all trials in each of the panels. This implies that 
a very small fraction are accepted. Numerically, we find that 
the $6$-, $8$-, and $10$-Gyr isochrones have the highest rate of acceptance: 
about $0.2\%$ for each (see Table~\ref{table:three}). To the extent that 
these do not overlap (which is partial), they would add constructively. 
Thus, these three isochrones contribute about $0.6\%$. 
The other two isochrones contribute negligibly.

We next focus on the ${\rm [Fe/H]} = +0.3$ isochrones. Again, the oldest three 
isochrones contribute the most. However, such old, very metal rich stars are 
very rare within a few kpc of the Sun. Hence, we ignore these. The two youngest 
isochrones together contribute $< 1\%$.

Lastly, we examine the solar-metallicity isochrones. These contribute 
$(1.4\%, 3.0\%, 3.7\%)$ for the $(6, 8, 10)$ Gyr isochrones, respectively. 
However, $10$-Gyr solar-metallicity stars are extremely rare, and $8$-Gyr 
solar-metallicity stars are fairly rare, so we make an overall estimate 
of $3\%$ for solar metallicity stars.  
We note that the two youngest isochrones contribute negligibly.

\begin{figure}[th]
\epsscale{0.9}
\plotone{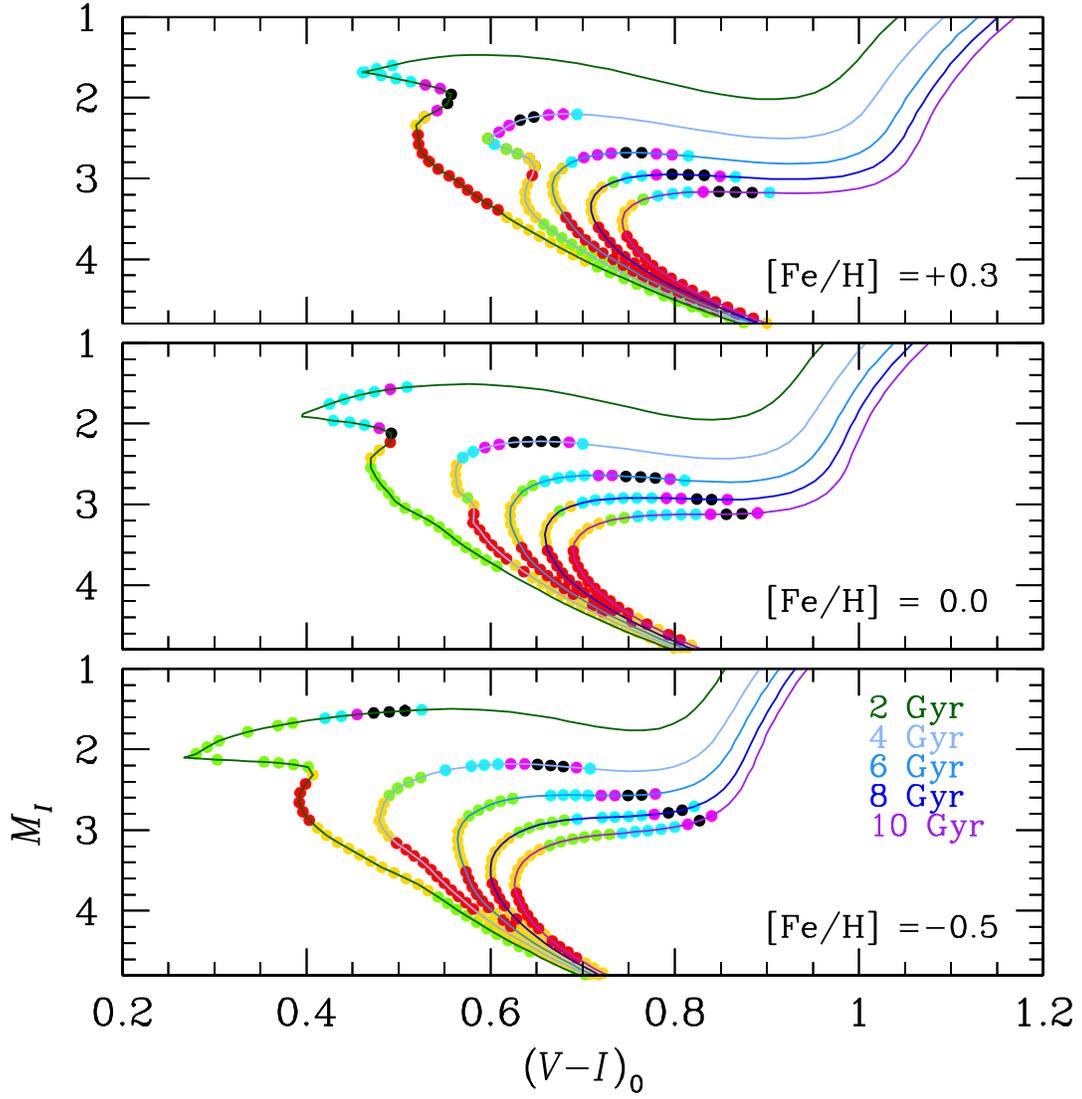}
\caption{\label{fig:eleven}
Positions of lens hosts and companions on $15$ model isochrones from the ``accepted'' events.
In each panel, the dots with (red, yellow, green) and (black, magenta, cyan) colors are 
the $(1, 2, 3)\sigma$ positions of the host and its companion, respectively.
}
\end{figure}

Next, we examine Figure~\ref{fig:eleven}, which shows where hosts (red,
yellow, green) and stellar companions (black, magenta, cyan) lie on the
theoretical isochrones for all $15$ isochrone cases. We note that only
``accepted'' events are shown. The most important feature of these
diagrams is that the stellar companion tracks are almost all confined to the
subgiant branch. This confirms the basic logic of the approach that we
outlined in the enumeration in Section~\ref{sec:five_2}, i.e.,
of considering the relative probability of systems that contain a
subgiant-branch star. Recall that if the stellar companion were on the
main-sequence for case (3), but it were on the subgiant branch for
case (2), then we would need to take account of the fact that
main-sequence stars are more common than subgiants.

Now, the companion is actually on the main sequence for the top two $2$-Gyr 
isochrones, and it is on the turnoff for the metal rich $4$-Gyr isochrone. 
However, recall from Figure~\ref{fig:ten} that the former contributes 
negligibly and the latter contributes $ < 1\%$. Even if this percentage 
were augmented by a factor $\sim 5$ due to slower evolution on the turnoff, 
its contribution would still be small.

Thus, considering that both ${\rm [Fe/H]} = 0.0$ and ${\rm [Fe/H]} = +0.3$ 
can contribute to case (1) as discussed in Section~\ref{sec:five_1}, 
while $ < 5\%$ of stellar populations at
these metallicities can contribute to cases (2)+(3), 
we estimate that from this quantitative analysis alone, the probability 
for cases (2)+(3) relative to case (1) is $p_a < 5\%$.

We now must take account of the fact that Classes (A), (C), (D), and (E)
can contribute to case (1), while Classes (E) and (F) can contribute
to cases (2)+(3). This contributes a relative probability of 
$p_b/(1-p_b) = (0.05\times0.84 + 0.01\times1.9)/(0.36+0.28+0.10+0.05) = 0.077$,
i.e., $p_b = 7\%$. Therefore, the ``total probability'' that the
host dominates the blend light is $(1 - p_a{\times}p_b) > 99.6\%$.

Finally, we ask why the quantitative analysis gave much more certainty 
(``$>99.6\%$'') that the host dominates the blended light than the 
qualitative analysis? The primary reason is that in the qualitative 
analysis, we implicitly assumed that, for most cases, there would be 
some isochrone that could provide the ``extra light'' from a 
turnoff/subgiant star that could be added to the host to make the 
observed blended light. However, Figure~\ref{fig:ten} shows that 
this is not the case.

\section{Discussion}
\label{sec:discuss}

We have shown that the bright, relatively blue $[(V-I), I] \sim (1.8, 15.8)$ 
blended light is very likely to be primarily due to the host. The blend, and 
thus almost certainly the host, can be basically characterized immediately 
from a medium-resolution spectrum taken on a ${\rm 4\,m}$, or even ${\rm 2\,m}$ class telescope. 
This would also provide a first epoch for the radial-velocity signatures of a putative stellar companion 
to the blend. Moreover, by taking a high-resolution spectrum on 
an ${\rm 8\,m}$ class telescope (similar to those obtained by \citealt{bensby13}), 
one could make a very detailed study of the chemical composition, age and mass of the blend/host.

Finally, future radial-velocity observations with ${\rm 30\,m}$ class telescopes 
could potentially detect and further characterize the planet. 
For example, let us assume that host and planet have $(M, m_{\rm p}, a_\perp) = (1.16\,M_\odot, 0.74\,M_{\rm J}, 4.5\,{\rm au})$, 
as in the example of the 6-Gyr, ${\rm [Fe/H]} = 0.0$ isochrone analyzed
in Section~\ref{sec:five_1}. Then, we may estimate a semi-major axis, $a = \sqrt{3/2}\,a_\perp = 5.5\,{\rm au}$, 
i.e., very similar to our own Jupiter. In this case, the period would be 
$P = 12\,$yr, and reflex velocity of the host would be $v = 8.5\,{\rm m\,s^{-1}}$. 
While the amplitude of this variation will be further reduced by $v\rightarrow v\sin i$,
it should still be measurable on ${\rm 30\,m}$ class telescopes. Because we already know $q$, 
these measurements would enable determination of the inclination angle $i$, 
in addition to the period $P$ and the eccentricity $e$, which are rarely 
if ever possible for microlensing planets.

OGLE-2018-BLG-1269Lb is the second microlensing planet with a bright blue host 
for which such spectroscopic studies on ${\rm 30\,m}$ telescopes will be possible. 
The first was OGLE-2018-BLG-0740b \citep{han19}, which also had a bright, 
blue blend due to a $\sim 1.0\,M_\odot$ host. In that case, the host 
was more than one magnitude fainter in the $I$ band, but just $0.65$ mag 
fainter in the $V$ band compared to OGLE-2018-BLG-1269Lb. On the other hand, 
the planet-host mass ratio $q$ was substantially larger, leading to an estimated 
reflex velocity $v$ that was $7.5$ times larger. Taking all these factors into 
account, OGLE-2018-BLG-0740Lb and OGLE-2018-BLG-1269Lb are comparably feasible 
for future radial-velocity studies\footnote{The first microlensing 
planets for which $30\,$m telescope radial-velocity studies were proposed 
were OGLE-2006-BLG-109Lb,c. At $M = 0.5\,M_\odot$, their host is much 
less massive, hence much redder and fainter than either of the
bright blue hosts discussed here. Nevertheless, \citet{bennett10} 
estimated that the host had $ H_{\rm L} = 17.2$ and so proposed that 
it would be possible to monitor it in the infrared with $30\,$m telescopes.}.

\acknowledgments
This research has made use of the KMTNet system operated by the Korea 
Astronomy and Space Science Institute (KASI) and the data were obtained at 
three host sites of CTIO in Chile, SAAO in South Africa, and SSO in Australia. 
AG was supported by JPL grant 1500811. 
Work by CH was supported by the grants of National Research Foundation of Korea (2017R1A4A1015178 and 2019R1A2C2085965).
The OGLE has received funding from the National Science Centre, Poland, grant MAESTRO 2014/14/A/ST9/00121 to A.U. 
The MOA project is supported by JSPS KAKENHI Grant Number JSPS24253004, JSPS26247023, JSPS23340064, JSPS15H00781, JP16H06287 and 19KK0082.

\begin{deluxetable}{lrrrr}
\tabletypesize{\footnotesize}
\tablecaption{Lensing Parameters\label{table:one}}
\tablewidth{0pt}
\tablehead{
\multicolumn{1}{l}{Parameters} &
\multicolumn{1}{c}{Local A} &
\multicolumn{3}{c}{Local B} \\
\multicolumn{1}{c}{} & 
\multicolumn{1}{c}{Standard} &
\multicolumn{1}{c}{Standard} &
\multicolumn{2}{c}{Orbit+Parallax} \\
\multicolumn{1}{c}{} & 
\multicolumn{1}{c}{} & 
\multicolumn{1}{c}{} & 
\multicolumn{1}{c}{$u_{0} > 0$} & 
\multicolumn{1}{c}{$u_{0} < 0$} 
}
\startdata
$\chi^2_{\rm tot}$/dof          &   22266.3/21841        &    22229.4/21841      &    22221.2/21837       &     22222.1/21837     \\
$t_0$ (${\rm HJD'}$)            & 8343.876$\pm$ 0.025    & 8343.849$\pm$ 0.025   & 8343.903$\pm$ 0.030    & 8343.895$\pm$ 0.029   \\
$u_0$                           &   0.141 $\pm$ 0.026    &   0.142 $\pm$ 0.023   &   0.144 $\pm$ 0.028    &  -0.143 $\pm$ 0.024   \\
$t_{\rm E}$ (days)              &  70.792 $\pm$ 1.064    &  70.584 $\pm$ 0.923   &  70.672 $\pm$ 1.661    &  69.597 $\pm$ 1.172   \\
$s$                             &   1.032 $\pm$ 0.019    &   1.126 $\pm$ 0.012   &   1.123 $\pm$ 0.032    &   1.124 $\pm$ 0.033   \\
$q$ ($10^{-4}$)                 &   5.940 $\pm$ 0.063    &   5.932 $\pm$ 0.066   &   5.753 $\pm$ 0.264    &   5.957 $\pm$ 0.230   \\
$\alpha$ (rad)                  &   1.888 $\pm$ 0.026    &   1.887 $\pm$ 0.026   &   1.887 $\pm$ 0.069    &  -1.891 $\pm$ 0.068   \\
$\rho$ ($10^{-4}$)              &   5.886 $\pm$ 0.097    &   5.917 $\pm$ 0.092   &   5.895 $\pm$ 0.177    &   5.941 $\pm$ 0.130   \\
$\pi_{{\rm E}, N}$              &   --                   &   --                  &   0.171 $\pm$ 0.150    &   0.114 $\pm$ 0.173   \\
$\pi_{{\rm E}, E}$ ($10^{-1}$)  &   --                   &   --                  &   0.086 $\pm$ 0.217    &   0.253 $\pm$ 0.107   \\
$ds/dt$ (yr$^{-1}$)             &   --                   &   --                  &  -0.287 $\pm$ 0.319    &  -0.219 $\pm$ 0.298   \\
$d\alpha/dt$ (yr$^{-1}$)        &   --                   &   --                  &   0.032 $\pm$ 0.554    &   0.205 $\pm$ 0.492   \\    
$f_{\rm s}$                     &   0.270 $\pm$ 0.006    &   0.270 $\pm$ 0.005   &   0.275 $\pm$ 0.007    &   0.273 $\pm$ 0.006   \\  
$f_{\rm b}$                     &   7.303 $\pm$ 0.006    &   7.304 $\pm$ 0.005   &   7.299 $\pm$ 0.007    &   7.301 $\pm$ 0.006   
\enddata 
\vspace{0.05cm}
\end{deluxetable}

\begin{deluxetable}{lrrr}
\tablecaption{Physical Parameters\label{table:two}}
\tablewidth{0pt}
\tablehead{
\multicolumn{1}{l}{Parameters} &
\multicolumn{1}{c}{$u_{0} > 0$} &
\multicolumn{1}{c}{$u_{0} < 0$} &
\multicolumn{1}{c}{Weighted} 
}
\startdata
$M_{1}$ $(M_{\odot})$                              &  $ 1.11_{-0.34}^{+0.68}$     &  $ 1.18_{-0.36}^{+0.78}$   &  $ 1.13_{-0.35}^{+0.72}$    \\ 
$M_{2}$ $(M_{\rm J})$                              &  $ 0.67_{-0.21}^{+0.41}$     &  $ 0.74_{-0.23}^{+0.49}$   &  $ 0.69_{-0.22}^{+0.44}$    \\ 
$a_{\bot}$ (au)                                    &  $ 4.51_{-1.15}^{+1.66}$     &  $ 4.75_{-1.21}^{+1.74}$   &  $ 4.61_{-1.17}^{+1.70}$    \\
$D_{\rm L}$ (kpc)                                  &  $ 2.51_{-0.61}^{+0.90}$     &  $ 2.64_{-0.64}^{+0.94}$   &  $ 2.56_{-0.62}^{+0.92}$    \\
$\pi_{{\rm E}, N}$                                 &  $0.187_{-0.071}^{+0.078}$   &  $0.174_{-0.071}^{+0.074}$ &  $0.183_{-0.071}^{+0.077}$  \\    
$\pi_{{\rm E}, E}$                                 &  $0.011_{-0.017}^{+0.014}$   &  $0.029_{-0.014}^{+0.011}$ &  $0.018_{-0.019}^{+0.014}$  \\  
$\mu_{{\rm rel, hel}}(N)$ $({\rm mas\,yr^{-1}})$   &  $ 8.01_{-0.65}^{+0.59}$     &   $7.91_{-0.68}^{+0.63}$   &   $7.97_{-0.66}^{+0.62}$    \\   
$\mu_{{\rm rel, hel}}(E)$ $({\rm mas\,yr^{-1}})$   &  $ 1.56_{-0.56}^{+0.57}$     &   $2.37_{-0.56}^{+0.63}$   &   $1.84_{-0.68}^{+0.81}$    \\   
$\phi_{\rm hel}$ (deg)                             &  $11.02_{-3.28}^{+2.89}$     &  $16.68_{-2.63}^{+2.68}$   &  $13.00_{-3.98}^{+4.14}$    \\
$(V-I)_{\rm L}$                                    &  $ 1.76_{-0.20}^{+0.44}$     &   $1.74_{-0.19}^{+0.41}$   &   $1.75_{-0.20}^{+0.44}$    \\
$I_{\rm L}$                                        &  $17.40_{-1.06}^{+1.15}$     &  $17.30_{-1.03}^{+1.15}$   &  $17.36_{-1.04}^{+1.16}$    \\
$P_{\rm lc}$                                       &  1.0                         &   0.64                     &  --                         \\
$P_{\rm tot}$                                      &  416913.4                    &   382315.5                 &  --                         \\
$P_{\rm net}$                                      &  416913.4                    &   244681.9                 &  --
\enddata 
\vspace{0.05cm}
\end{deluxetable}

\begin{deluxetable}{lrrrr}
\tablecaption{Rate of Acceptance\label{table:three}}
\tablewidth{0pt}
\tablehead{
\multicolumn{1}{l}{age (Gyr)} &
\multicolumn{1}{c}{${\rm [Fe/H]} = -0.5$} & 
\multicolumn{1}{c}{${\rm [Fe/H]} = 0.0$} & 
\multicolumn{1}{c}{${\rm [Fe/H]} = +0.3$}
}
\startdata
$2$         &  $3.56\times10^{-6}$  &  $1.15\times10^{-4}$  &  $2.17\times10^{-3}$  \\ 
$4$         &  $4.32\times10^{-4}$  &  $1.86\times10^{-3}$  &  $5.97\times10^{-3}$  \\ 
$6$         &  $1.67\times10^{-3}$  &  $1.43\times10^{-2}$  &  $2.71\times10^{-2}$  \\ 
$8$         &  $2.24\times10^{-3}$  &  $2.97\times10^{-2}$  &  $5.69\times10^{-2}$  \\ 
$10$        &  $2.37\times10^{-3}$  &  $3.72\times10^{-2}$  &  $8.19\times10^{-2}$ 
\enddata 
\vspace{0.05cm}
\end{deluxetable}

\begin{thebibliography}{99}

\bibitem[Alard \& Lupton(1998)]{alard98}
Alard, C., \& Lupton, Robert H. 1998, \apj, 503, 325

\bibitem[Albrow et al.(2009)]{albrow09}
Albrow, M. D., Horne, K., Bramich, D. M., et al. 2009, \mnras, 397, 2099

\bibitem[Alcock et al.(2001)]{alcock01}
Alcock, C., Allsman, R. A., Alves, D. R., et al. 2001, \nat, 414, 617

\bibitem[Batista et al.(2015)]{batista15}
Batista, V., Beaulieu, J.-P., Bennett, D. P., et al. 2015, \apj, 808, 170

\bibitem[Batista et al.(2011)]{batista11}
Batista, V., Gould, A., Dieters, S., et al. 2011, \aap, 529, 102

\bibitem[Beaulieu et al.(2016)]{beaulieu16}
Beaulieu, J.-P., Bennett, D. P., Batista, V., et al. 2016, \apj, 824, 83

\bibitem[Bennett et al.(2015)]{bennett15}
Bennett, D. P., Bhattacharya, A., Anderson, J., et al. 2015, \apj, 808, 169

\bibitem[Bennett et al.(2020)]{bennett20}
Bennett, D. P., Bhattacharya, A., Beaulieu, J. P.,et al. 2020, \aj, 159, 68

\bibitem[Bennett et al.(2010)]{bennett10}
Bennett, D. P., Rhie, S. H., Nikolaev, S., et al. 2010, \apj, 713, 837

\bibitem[Bensby et al.(2013)]{bensby13}
Bensby, T., Yee, J. C., Feltzing, S., et al. 2013, \aap, 549, 147

\bibitem[Bessell \& Brett(1988)]{bessell88}
Bessell, M. S., \& Brett, J. M. 1988, \pasp, 100, 1134

\bibitem[Bhattacharya et al.(2018)]{bhattacharya18}
Bhattacharya, A., Beaulieu, J.-P., Bennett, D.P., et al. 2018, \aj, 156, 289

\bibitem[Bond et al.(2001)]{bond01}
Bond, I. A., Abe, F., Dodd, R. J., et al. 2001,\mnras, 327, 868

\bibitem[Dominik(1998)]{dominik98}
Dominik, M. 1998, \aap, 329, 361

\bibitem[Dong et al.(2009)]{dong09}
Dong, S., Gould, A., Udalski, A., et al. 2009, \apj, 695, 970

\bibitem[Duquennoy \& Mayor(1991)]{duquennoy91}
Duquennoy, A., \& Mayor, M. 1991, \aap, 248, 485

\bibitem[Gaia Collaboration et al.(2018)]{gaia18}
Gaia Collaboration, Brown, A. G. A., Vallenari, A., et al. 2018, \aap, 616, 1

\bibitem[Gaia Collaboration et al.(2016)]{gaia16}
Gaia Collaboration, Prusti, T., de Bruijne, J. H. J., et al. 2016, \aap, 595, A1

\bibitem[Gaudi et al.(2008)]{gaudi08}
Gaudi, B. S., Bennett, D. P., Udalski, A., et al. 2008, Sci, 319, 927

\bibitem[Gould(1992)]{gould92}
Gould, A. 1992, \apj, 392, 442

\bibitem[Gould(2000)]{gould00} 
Gould, A. 2000, \apj, 542, 785

\bibitem[Griest et al.(1991)]{griest91} 
Griest, K. et al. 1991, \apj, 372, L79

\bibitem[Han et al.(2018)]{han18}
Han, C., Jung, Y. K., Udalski, A., et al. 2018, \apj, 867, 136

\bibitem[Han et al.(2019)]{han19}
Han, C., Yee, J. C., Udalski, A., et al. 2019, \aj, 158, 102

\bibitem[Hirao et al.(2020)]{hirao20} 
Hirao, Y., Bennett, D.P., Ryu, Y.-H., et al., 2020, in press, arXiv:2004.09067

\bibitem[Jung et al.(2019)]{jung19}
Jung, Y. K., Gould, A., Zang, W., et al. 2019, \aj, 157, 72

\bibitem[Jung et al.(2018)]{jung18}
Jung, Y. K., Udalski, A., Gould, A., et al. 2018, \aj, 155, 219

\bibitem[Jung et al.(2015)]{jung15}
Jung, Y. K., Udalski, A., Sumi, T., et al. 2015, \apj, 798, 123

\bibitem[Kervella et al.(2004)]{kervella04}
Kervella, P., Bersier, D., Mourard, D., et al. 2004, \aap, 428, 587

\bibitem[Kim et al.(2018)]{kim18}
Kim, D.-J., Kim, H.-W., Hwang, K.-H., et al. 2018, \aj, 155, 76

\bibitem[Kim et al.(2016)]{kim16}
Kim, S.-L., Lee, C.-U., Park, B.-G., et al. 2016, JKAS, 49, 37

\bibitem[Koz{\l}owski et al.(2007)]{kozlowski07}
Koz{\l}owski, S., Wo\'{z}niak, P. R., Mao, S., \& Wood, A. 2007, \apj, 671, 420

\bibitem[Luri et al.(2018)]{luri18}
Luri, X., Brown, A. G. A., Sarro, L. M., et al. 2018, \aap, 616, A9

\bibitem[Nataf et al.(2013)]{nataf13}
Nataf, D. M., Gould, A., Fouqu\'{e}, P., et al. 2013, \apj, 769, 88

\bibitem[Paczy\'{n}ski(1986)]{paczynski86}
Paczy\'{n}ski, B. 1986, \apj, 304, 1

\bibitem[Paczy\'nski(1991)]{paczynski91} 
Paczy\'{n}ski, B. 1991, \apj, 371, L63

\bibitem[Ryu et al.(2019)]{kb181292} Ryu, Y.-H., Navarro, M.G., Gould, A. et al. 2019, \aj, 159, 58

\bibitem[Skowron et al.(2011)]{skowron11}
Skowron, J., Udalski, A., Gould, A., et al. 2011, \apj, 738, 87

\bibitem[Spada et al.(2017)]{spada17}
Spada, F., Demarque, P., Kim, Y.-C., et al. 2017, \apj, 838, 161

\bibitem[Sumi et al.(2003)]{sumi03}
Sumi, T., Abe, F., Bond, I. A., et al. 2003, \apj, 591, 20

\bibitem[Szyma\'{n}ski et al.(2011)]{szymanski11}
Szyma\'{n}ski, M. K., Udalski, A., Soszy\'{n}ski, I., et al. 2011, AcA, 61, 83

\bibitem[Tomaney \& Crotts(1996)]{tomaney96}
Tomaney, A.B. \& Crotts, A.P.S. 1996, \aj, 112, 2872

\bibitem[Udalski(2003)]{udalski03}
Udalski, A. 2003, Acta Astron., 53, 291

\bibitem[Udalski et al.(2015)]{udalski15}
Udalski, A., Szyma\'{n}ski, M. K., \& Szyma\'{n}ski, G. 2015, AcA, 65, 1

\bibitem[Vandorou et al.(2019)]{vandorou19}
Vandorou, A., Bennett, D. P., Beaulieu, J.-P., et al. 2019, arXiv:1909.04444

\bibitem[Wo\'{z}niak(2000)]{wozniak00}
Wo\'{z}niak, P. R. 2000, AcA, 50, 42

\bibitem[Yee et al.(2016)]{yee16}
Yee, J. C., Johnson, J. A., Skowron, J., et al. 2016, \apj, 821, 121

\bibitem[Yoo et al.(2004)]{yoo04}
Yoo, J., DePoy, D. L., Gal-Yam, A., et al. 2004, \apj, 603, 139

\bibitem[Zinn et al.(2019)]{zinn19}
Zinn, J. C., Pinsonneault, M. H., Huber, D., \& Stello, D. 2019, \apj, 878, 136




\end{thebibliography}
\end{document}